\documentclass[journal]{IEEEtran}
\usepackage{mathtools, amsthm, amsmath,amssymb,amsfonts, dsfont}
\usepackage{textcomp, acronym, cite, url, comment, footnote}
\usepackage{graphicx, tikz, subcaption, xcolor, enumitem, xspace, diagbox, fancyhdr, multicol, multirow, adjustbox}
\usepackage[bookmarks,colorlinks]{hyperref} 
\usepackage[linesnumbered,ruled,vlined]{algorithm2e}
\let\oldnl\nl
\newcommand{\nonl}{\renewcommand{\nl}{\let\nl\oldnl}} 
\setlist[description]{font=\normalfont}
\def\BibTeX{{\rm B\kern-.05em{\sc i\kern-.025em b}\kern-.08em 
T\kern-.1667em\lower.7ex\hbox{E}\kern-.125emX}}

\acrodef{ml}[ML]{machine learning}
\acrodef{fl}[FL]{federated learning}
\acrodef{dnn}[DNN]{deep neural network}
\acrodef{dq}[DQ]{dithered quantization}
\acrodef{sdq}[SDQ]{subtractive dithered quantization}
\acrodef{snr}[SNR]{signal-to-noise ratio}
\acrodef{mlp}[MLP]{multi-layer perceptron}
\acrodef{cnn}[CNN]{convolutional neural network}
\acrodef{jopeq}[JoPEQ]{joint privacy enhancement and quantization} 
\acrodef{mse}[MSE]{mean squared error} 
\acrodef{sgd}[SGD]{stochastic gradient descent}
\acrodef{fa}[FedAvg]{federated averaging} 
\acrodef{name}[OLALa]{Online Learned Adaptive Lattices} 

\newcommand{\sample}{i^u_t}

\newcommand{\myVec}[1]{{\boldsymbol{#1}}}

\newcommand{\cD}{\mathcal{D}}

\newcommand{\cL}{\mathcal{L}}

\newcommand{\cP}{\mathcal{P}}
\newcommand{\cO}{\mathcal{O}}
\newcommand{\cI}{\mathcal{I}}
\newcommand{\cX}{\mathcal{X}}

\newcommand{\cG}{\mathcal{G}}
\newcommand{\cF}{\mathcal{F}}

\newcommand{\R}{\mathbb{R}}

\newcommand{\Z}{\mathbb{Z}}

\newcommand{\E}{\mathds{E}}

\newcommand{\vw}{\boldsymbol{w}}
\newcommand{\vG}{\boldsymbol{G}}
\newcommand{\vs}{\boldsymbol{s}}
\newcommand{\vx}{\boldsymbol{x}}

\newcommand{\vz}{\boldsymbol{z}}
\newcommand{\vl}{\boldsymbol{l}}
\newcommand{\vh}{\boldsymbol{h}}

\newcommand{\ve}{\boldsymbol{e}}
\newcommand{\vd}{\boldsymbol{d}}
\newcommand{\vg}{\boldsymbol{g}}

\newcommand{\vA}{\boldsymbol{A}}

\newcommand{\Var}{\mathrm{Var}}
\DeclareMathOperator*{\argmin}{arg\,min}

\newcommand{\lsgd}{local-\ac{sgd}\xspace}

\newtheorem{theorem}{Theorem}[section]
\newtheorem{definition}[theorem]{Definition}

\newtheorem{lemma}{Lemma}

\usepackage[all=normal,paragraphs=tight,floats=normal,mathspacing=normal,wordspacing=tight,charwidths=tight,mathdisplays=normal,leading=normal]{savetrees}

 \usepackage{algorithmic}
\usepackage{multirow}
\usepackage{booktabs}
\usepackage{graphicx}
\usepackage{colortbl}
\usepackage{array}

\begin{document}

\title{OLALa: Online Learned Adaptive Lattice Codes \\for Heterogeneous Federated Learning}

\author{Natalie Lang$^*$, Maya Simhi$^*$, and
Nir Shlezinger~\thanks{$^*$Equal contribution. The preliminary findings of this work were presented in the 2024 IEEE International Conference on Acoustics, Speech, and Signal Processing (ICASSP) as the paper \cite{lang2024data}.  The authors are with the School of ECE, Ben-Gurion University of the Negev Be’er-Sheva, Israel. 
Emails: \texttt{\{langn,mayasimh\}@post.bgu.ac.il, nirshl@bgu.ac.il}}}

\maketitle

\begin{abstract}
\Ac{fl} enables collaborative training across distributed clients without sharing raw data, often at the cost of substantial communication overhead induced by transmitting high-dimensional model updates. This overhead can be alleviated by having the clients quantize their model updates, with dithered lattice quantizers identified as an attractive scheme due to its structural simplicity and convergence-preserving properties. However, existing lattice-based FL schemes typically rely on a fixed quantization rule, which is suboptimal in heterogeneous and dynamic environments where the model updates distribution varies across users and training rounds. In this work, we propose {\em {\bf O}nline {\bf L}earned {\bf A}daptive {\bf La}ttices (OLALa)}, a  heterogeneous \ac{fl} framework where each client can adjust its quantizer online using lightweight local computations. We first derive convergence guarantees for \ac{fl} with non-fixed lattice quantizers and show that proper lattice adaptation can tighten the convergence bound. Then, we design an online learning algorithm that enables clients to tune their quantizers throughout the \ac{fl} process while exchanging only a compact set of quantization parameters. Numerical experiments demonstrate that \acs{name} consistently improves learning performance under various quantization rates, outperforming conventional fixed-codebook and non-adaptive schemes.
\end{abstract}

\acresetall

\section{Introduction}\label{sec:intro}
\IEEEPARstart{T}{he}  machine learning framework of \ac{fl} has multiple clients collaboratively train a global model under the orchestration of a central server, while keeping their local data private~\cite{mcmahan2017communication}. By transmitting only model updates rather than raw data, \ac{fl} enables the utilization of distributed datasets without requiring private data to be shared~\cite{abdulrahman2020survey,wen2023survey, kairouz2021advances}.

One of the principal challenges of \ac{fl}, that is largely absent in centralized learning setups, is the substantial communication overhead stemming from the frequent exchange of high-dimensional model updates between the server and the distributed clients~\cite{kairouz2021advances}. This communication bottleneck is particularly critical in bandwidth-limited or wireless environments, where the transmission cost can dominate the training process~\cite{gafni2021federated}. As a result, a broad range of techniques have been proposed to mitigate communication load, including client selection strategies~\cite{chen2021communication,peleg2025pause}, update sparsification~\cite{han2020adaptive, aji2017sparse, alistarh2018convergence}, and over-the-air aggregation schemes~\cite{amiri2020machine, sery2021over,yang2020federated}. A prominent and widely adopted approach is {\em model compression}, wherein each client transmits a compressed representation of its local update~\cite{chen2024learned}. In particular, lossy compression methods, and notably quantization~\cite{alistarh2017qsgd,reisizadeh2020fedpaq,bernstein2018signsgd,shlezinger2020uveqfed, horvath2019natural}, have demonstrated significant effectiveness in reducing communication costs while maintaining acceptable learning performance~\cite{wang2021quantized}.

Generally speaking, quantization can be represented as a codebook of digital representations~\cite{gray1998quantization}. Such codebooks can  be tuned  from data using classic clustering-type methods~\cite{linde1980algorithm,kohonen2001learning}, with more recent techniques employing deep learning~\cite{van2017neural,fishel2025remote}. In practice, one is often interested in utilizing structured mappings, and particularly uniform quantizers and their vector general form of {\em lattice quantizers}, as opposed to arbitrary codebooks~\cite{shlezinger2018hardware}. Such structured quantizers are simple, support nested implementations~\cite{polyanskiy2014lecture}, and, when combined with dithering~\cite{gray1993dithered,lipshitz1992quantization}, can eliminate dependence between the distortion and the signal~\cite{zamir1992universal}. Such forms of dithered quantization are considered to be highly suitable for model update compression in \ac{fl}, due their ability to preserve convergence of the training procedure~\cite{alistarh2017qsgd, shlezinger2020uveqfed}, while potentially enhancing privacy preservation~\cite{lang2023joint, shahmiri2024communication, lang2023compressed}.

The mapping of a lattice quantizer is determined by its {\em generator matrix} and the associated distance metric~\cite{zamir1996lattice}. Traditional designs of dithered lattice quantizers leverage their tractable rate-distortion behavior under the assumption that the quantizer is not overloaded. This property allows for analytical bounds on the quantization error characterized in terms of the lattice second-order moment to guide the design~\cite{gersho1979asymptotically}, either through closed-form expressions in special cases~\cite{lyu2022better, agrell2023best} or by iterative optimization procedures~\cite{agrell1998optimization, agrell2025optimization}. Alternatively, our preliminary study~\cite{lang2024data} showed that  deep learning tools can learn lattice quantizers that directly minimize distortion without relying on non-overload analytical bounds, by converting the  quantizer into a trainable discriminative model~\cite{shlezinger2022discriminative}. In the context of \ac{fl}, these strategies result in a single, fixed quantization rule that is uniformly applied across all clients and throughout all communication rounds. However, due to the inherent statistical (data distribution) heterogeneity of \ac{fl}~\cite{ye2023heterogeneous}, combined with the evolving nature of model updates over time, the distribution of the weights to be quantized can vary significantly across clients and iterations. Moreover, these variations are often difficult to predict or model in advance, indicating that a one-size-fits-all lattice design may be suboptimal for \ac{fl}. 

In this work, we study heterogeneous \ac{fl} under user- and time-varying adaptive quantization, motivated by the need to support dynamic and non-stationary weight distributions across clients and communication rounds. We introduce {\em \ac{name}}, a framework that enables locally adaptive dithered lattice quantization tailored to the evolving characteristics of each client’s updates. A key insight underlying \ac{name} is that dithered lattice quantizers are particularly well-suited to this setting, as their overall behavior is governed  by the compact generator matrix, whose dimensionality is independent of the quantization rate. This structural property makes them ideal for online adaptation: clients can efficiently adjust their quantizers on-the-fly based on local statistics and communicate the updated parameters to the server with negligible communication overhead, thus not being restricted to pre-determined compression rule as in alternative forms of learned local mappings~\cite{chen2024learned}. 

To motivate the design of \ac{name}, we begin by providing a theoretical analysis of the convergence behavior of heterogeneous \ac{fl} when employing non-identical, adaptive lattice quantizers. Under standard assumptions on the learning model and data distribution~\cite{li2019convergence}, we rigorously prove that the training process converges to the minimizer of the empirical risk, and that when operating with a fixed compression bit budget, the convergence bound can be tightened through appropriate adaptive lattice design.

Building on this insight, we develop an online learning algorithm based on model-based deep learning principles~\cite{shlezinger2022model}, which enables each client to dynamically adapt its lattice quantizer throughout the \ac{fl} process. Our approach draws insights from~\cite{ulyanov2018deep}, utilizing dedicated fixed-input \ac{dnn} during training to learn how to represent the generator matrix, combined with a novel casting of the quantizer mapping that enables gradient-based learning. The resulting \ac{name} learns user- and time-specific lattices using lightweight local computations and communicates only a compact set of quantizer metadata, thereby maintaining a low communication footprint. Our experiments demonstrate that \ac{name} effectively reduces the quantization distortion, i.e., improving its associated \ac{mse}, and yields superior learning performance across a wide range of quantization rates.

The rest of this paper is organized as follows: Section~\ref{sec:system_model_and_preliminaries}  reviews preliminaries in \ac{fl} and lattice quantization. Section~\ref{sec:method} introduces the \ac{name} framework, motivates using adaptive lattices via a theoretical convergence analysis, and details the proposed online learning scheme. We numerically evaluate \ac{name} in Section~\ref{sec:experiments}, with concluding  remarks provided in Section~\ref{sec:conclusions}. 

Throughout this paper, we use boldface lower-case letters for vectors, e.g., $\vx$. We use calligraphic letters for sets, e.g., $\cX$, with $|\cX|$ being its cardinality. The stochastic expectation, variance, inner product, and $\ell_2$ norm are denoted by $\E[\cdot]$, $\Var(\cdot)$, $\langle\cdot;\cdot\rangle$, and $\|\cdot\|$, respectively; while $\R$ is the set of real numbers.

\section{System Model and Preliminaries}\label{sec:system_model_and_preliminaries}
In this section, we set the ground for the derivation of \ac{name}. We commence by presenting the system model of \ac{fl} in Subsection~\ref{subsec:fl}. Then,  we provide a description of lattice-based compression in Subsection~\ref{subsec:lattice_coding}. 

\subsection{Federated Learning}\label{subsec:fl}
The \ac{fl} paradigm \cite{mcmahan2017communication} constitutes a central server training a parameterized model $\vw \in \R^m$, utilizing data stored at a group of $U$ users indexed by $u \in \{1, ..., U\}$. Unlike traditional centralized learning, their corresponding datasets, denoted $\cD_1, \cD_2, \ldots, \cD_U$, cannot be transferred to the server due to privacy or communication restrictions.
Letting $F_u(\vw)$ denote the empirical risk function of the $u$th user, \ac{fl} aims to find the model parameters  that minimize the averaged empirical risk across all users, i.e.,
\begin{equation} \label{eq:optimal_fl_model}
\vw_{\rm opt} = \arg\min_{\vw} \left\{ F(\vw) \triangleq \frac{1}{U} \sum_{u=1}^{U} F_u(\vw) \right\}.
\end{equation}
In \eqref{eq:optimal_fl_model}, it is  assumed (for simplicity) that the local datasets are  of the same cardinality, and that all devices participate in training. 

\ac{fl} follows an iterative procedure operated in rounds~\cite{kairouz2021advances}. In the $t$th round, the server transmits the current global model $\vw_t$ to the clients, who each updates the model using its local dataset and computational resources. 
These updates are then sent back to the server, which aggregates them to form the new global model. 
 \ac{fl} typically involves training at the devices via variants of \lsgd~\cite{stich2018local}, which, in its simplest form, updates the weights via  
\begin{align}\label{eq:local_sgd}
    \vw^u_{t+1} \xleftarrow{}  \vw_t -\eta_t \nabla F_u\left(\vw_t;\sample\right);
\end{align}
where $\sample$ is the data sample index chosen uniformly from $\cD_u$, and $\eta_t$ is the learning rate. In general, the client can replace \eqref{eq:local_sgd} with multiple local iterations, and convey the model update $\vh_{t}^u :=  \vw^u_{t+1} - \vw_t$.
The model updates are aggregated by the server, most commonly using \ac{fa}~\cite{mcmahan2017communication}, i.e., 
\begin{equation} \label{eq:FedAvg_update_rule}
\vw_{t+1} \triangleq \frac{1}{U} \sum_{u=1}^{U} \vw^u_{t+1}= \vw_{t} 
+ \frac{1}{U} \sum_{u=1}^{U} \vh_{t}^u.
\end{equation}

The above conventional \ac{fl} formulation gives rise to several key challenges, that are not present in centralized learning~\cite{gafni2021federated}. For once, the distribution of the datasets often varies between the users. This form of {\em statistical heterogeneity} indicates that the distribution of the local updates $\vh_{t}^u$ can vary considerably between different users~\cite{ye2023heterogeneous}. Moreover, broadcasting the local update of each single edge client results in a frequent exchange of high-dimensional parameter vectors to enable aggregation via~\eqref{eq:FedAvg_update_rule}. This {\em communication overhead} can possibly load the communication network. The latter is often tackled by integration of quantization techniques, discussed next.

\subsection{Lattice Coding}\label{subsec:lattice_coding}
Quantization refers to the representation of continuous-valued signals via finite-bit representations~\cite{gray1998quantization}. The tradeoff between the quantization rate and the distortion induced by the quantization operation can be improved by jointly quantizing multiple samples via {\em vector quantization}~\cite{gray1984vector}.
A leading approach to implement vector quantizers is based on lattice quantization~\cite{zamir1992universal}:
\begin{definition}[Lattice Quantizer]\label{def:lattice_quantizer}
A lattice quantizer of dimension $L \in \Z^+$ and  generator matrix $\vG\ \in \R^{L\times L}$ maps a continuous-valued vector $\vx\in \R^L$ into a discrete representation $Q_\cL(\vx)$ by selecting the nearest point in the (Voronoi-cells) lattice $\cL \triangleq \{ \vG\vl: \vl\in\Z^L\}$ , i.e., 
\begin{equation}
\label{eqn:latticeQuant}
    Q_\cL(\vx) = \argmin_{\vz \in \cL}\|\vx-\vz\|. 
\end{equation}
To apply $Q_\cL$ to a vector $\vx \in \R^{ML}$, it is divided into ${[\vx_1,\dots,\vx_M]}^T$, and each sub-vector is quantized separately.
\end{definition}
It is noted that for $L=1$, $Q_\cL(\cdot)$ in~\eqref{eqn:latticeQuant} specializes scalar uniform quantization. While quantizing via \eqref{eqn:latticeQuant} formally requires exhaustive search, it can be approached using computationally efficient methods~\cite{agrell2002closest}.  

\begin{figure}
  \centering
    \includegraphics[width=0.24\textwidth]{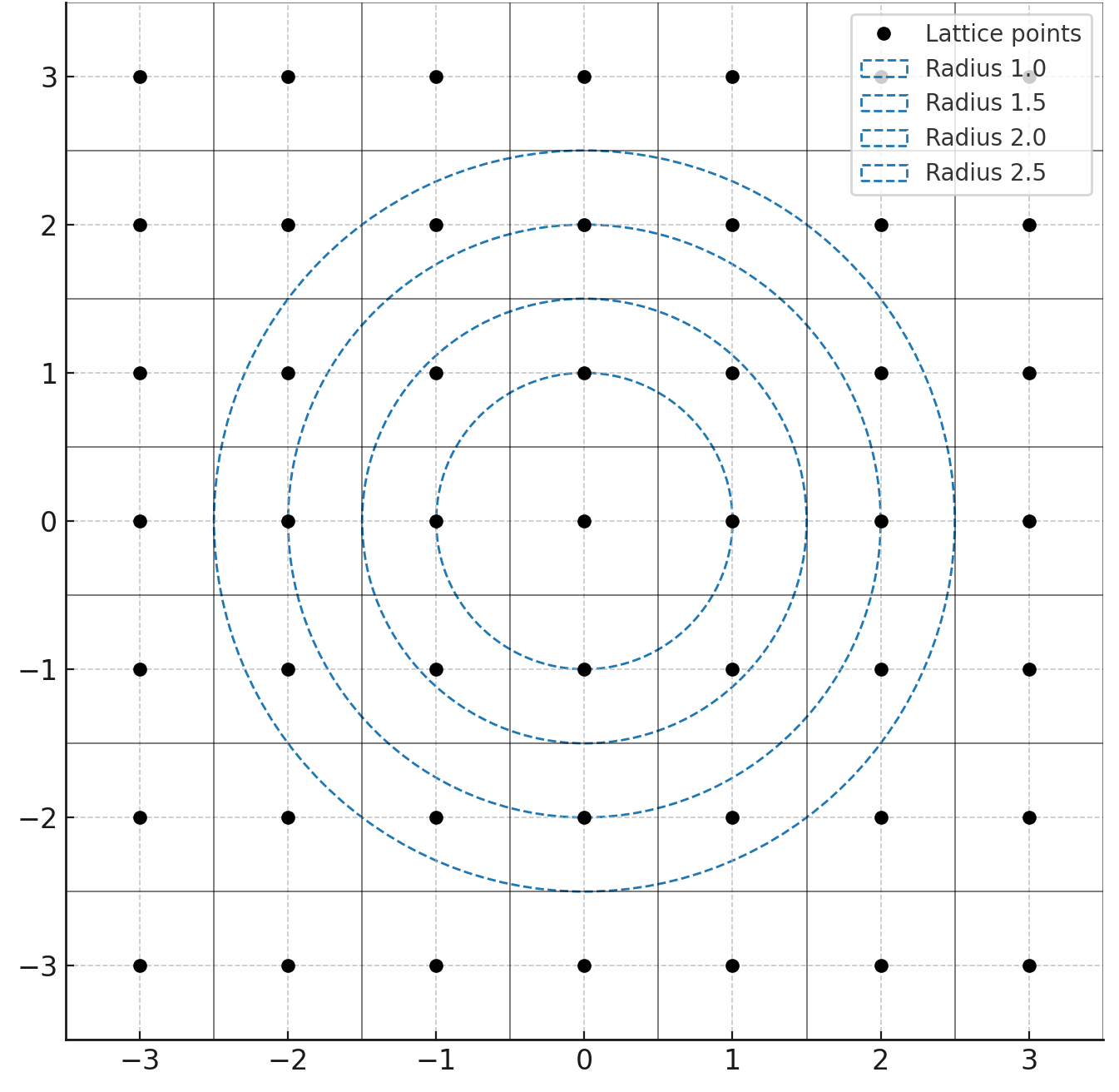}
    \includegraphics[width=0.24\textwidth]{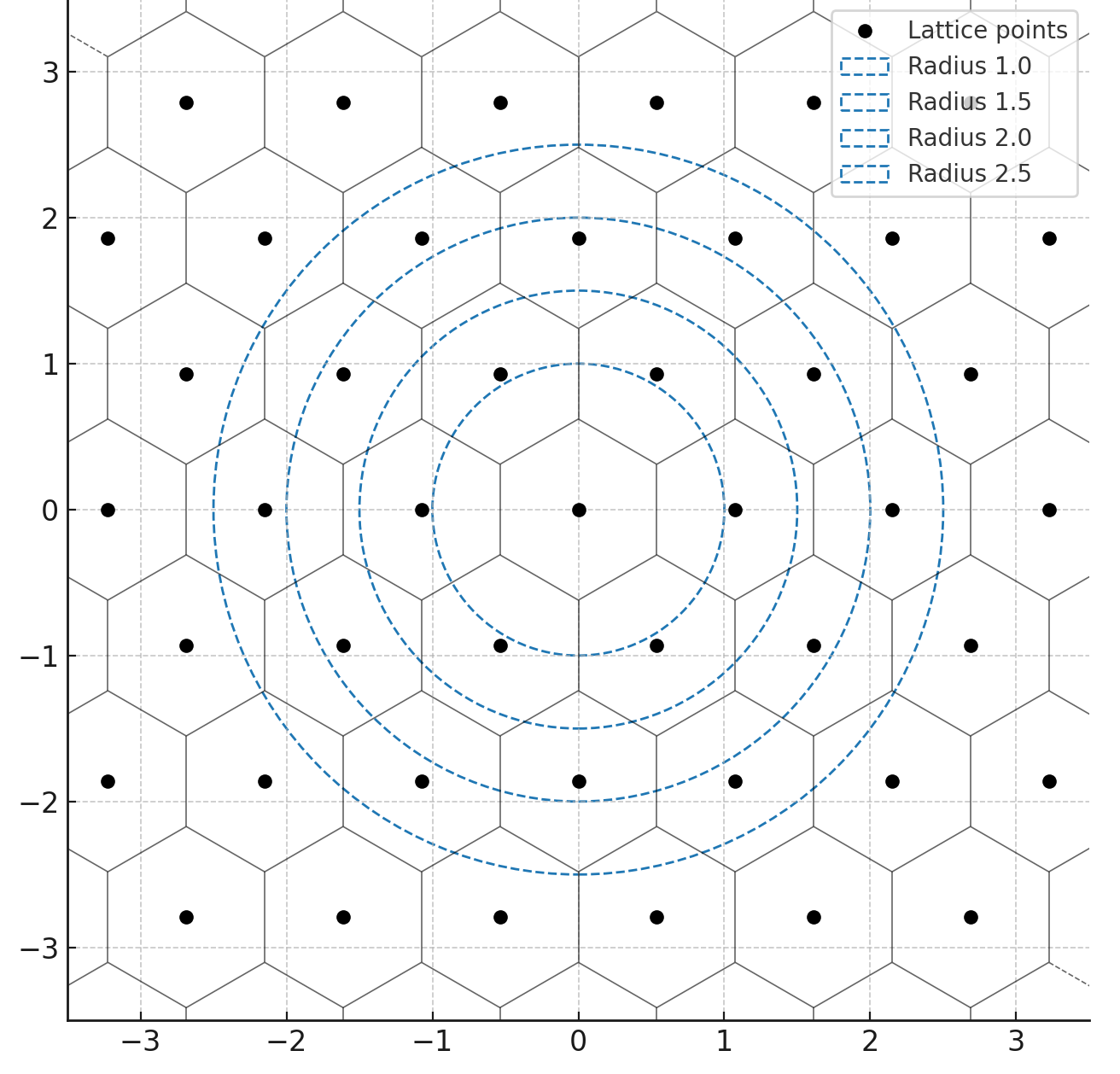}
  \caption{Example: truncated lattices structures.  
    \textcolor{black}{$\bullet$}'s mark lattice points (codewords); concentric circles 
    represent the radii (i.e., $\gamma$ values).
    The left quantizer has the identity as its generator matrix, while the right has the hexagonal generator matrix scaled to have unit determinant.}  
  \label{fig:lattices_example}
\end{figure}

A lattice $\cL$ partitions $\R^L$ into cells centered around the lattice points, where the basic cell is 
\begin{equation}\label{eq:basic_latice_cell}   
\cP_0=\{\vx:Q_\cL(\vx)=\boldsymbol 0\}. 
\end{equation}
The number of lattice points in $\cL$ is countable but infinite. Thus, to obtain a finite-bit representation, it is common to restrict $\cL$ to include only points in a given sphere of radius $\gamma$, $\cL_\gamma$, i.e.,
\begin{equation}\label{eqn:FinLattice}
    \cL_{\gamma}\triangleq\{\vz \in \cL: \|\vz\| \leq \gamma\}.
\end{equation}
The number of lattice points, $|\cL_\gamma|$, dictates the number of bits per sample, via the quantization rate $R\triangleq\frac{1}{L}\log_2(|\cL_\gamma|)$. 
An event in which the input to the lattice quantizer does not reside in this sphere is referred to as {\em overloading}, from which quantizers are typically designed to avoid \cite{gray1998quantization}. Fig.~\ref{fig:lattices_example} demonstrates different truncated lattices configurations in the $2$D plane, i.e., $L=2$.

Lattice quantization yields a distortion term  $\ve \triangleq Q_{\cL}(\vx) - \vx$ that is deterministically determined by $\vx$. To be made stochastic, it is often combined with {\em probabilistic quantization}, and particularly with \ac{sdq}~\cite{lipshitz1992quantization, gray1993dithered}:
\begin{definition}[\ac{sdq}]\label{def:SDQ} 
Let $\vd$ be drawn uniformly from $\cP_0$ in~\eqref{eq:basic_latice_cell}.
The  \acs{sdq} of $\vx \in \R^L$ with lattice $\cL$ is given by 
\begin{align}
    Q^{\rm SDQ}_{\cL}(\vx)= Q_\cL(\vx+\vd)-\vd.
\end{align}
\end{definition}
A key property of \ac{sdq} stems from the fact that its resulting distortion can be made independent of the quantized input, as stated in \cite[Thm. 1]{gray1993dithered}, recalled below as Theorem~\ref{thm:SDQ}:
\begin{theorem}\label{thm:SDQ}
Consider a quantizer induced by a $\gamma$-supported lattice $\cL_\gamma(\vG)$, with input vectors $\{\vx_i\}$ and  dither vectors $\{\vd_i\}$, where each  $\vd_i$ is uniformly distributed over the basic cell $\cP_0$ defined in~\eqref{eq:basic_latice_cell}. Then, if the quantizer is not over-loaded, i.e., $\forall i,\ \Pr(|\vx_i + \vd_i| < \gamma) = 1$,
the \ac{sdq} errors, defined as 
\begin{align}\label{eq:sdq_error}
    \ve_i^{\rm SDQ} &:=  Q^{\rm SDQ}_{\cL_\gamma(\vG)}(\vx_i) - \vx_i \notag\\
    &=\argmin_{\vl\in\cL_\gamma(\vG)} lang2024stragglers\|\vx_i+\vd_i - \vl\| -\vd_i - \vx_i \quad \forall i,
\end{align}
are mutually independent of $\{\vx_i\}$; and obey an i.i.d uniform distribution over the basic lattice cell. 
That is, $\E\left[\ve_i^{\rm SDQ}\right]=0$ and 
\begin{align}\label{eq:sdq_moments} 
\Var\left(\ve_i^{\rm SDQ}\right)=
\E\left[\left\|\ve_i^{\rm SDQ}\right\|^2\right]
=\sigma^2_{\rm SDQ}(\cL_\gamma(\vG)) \quad \forall i,
\end{align}
where $\sigma^2_{\rm SDQ}(\cL_\gamma(\vG))$ is the \ac{mse} per
dimension~\cite{zamir1996lattice},
\begin{equation}\label{eq:sigma2_sdq}
    \sigma^2_{\rm SDQ}(\cL_\gamma(\vG)) := \frac{1}{L}\cdot\frac{\int_{\cP_0} \|\vx\|^2 \, d\vx}{\text{Vol}(\cP_0)};
\end{equation}
and $\text{Vol}(\cP_0)$ is the volume of $\cP_0$. 
\end{theorem}

Theorem~\ref{thm:SDQ} indicates that when  the quantizer is not {overloaded}, the distortion induced by \ac{sdq} can be  modeled as white noise uniformly distributed over $\cP_0$. 
It also implies that, for any input vector residing within the support of the truncated-lattice, \ac{sdq} with multivariate lattice quantization yields distortion with lesser variance compared to uniform scalar quantizers with the same number of bits per sample~\cite{zamir1996lattice}. 



    

\section{Online Adaptive Lattice Learning}
\label{sec:method}
In this section we introduce the proposed \ac{name} framework. We  start in Subsection~\ref{ssec:FL_Adaptive} by explaining the operation of \ac{fl} with lattice quantizers in which the lattices are allowed to vary, without specifying how they are being set. Then, we theoretically analyze the convergence profile of the resulting \ac{fl} framework in Subsection~\ref{subsec:analysis}, indicating the theoretical potential gains of adapting the lattices between users and rounds in heterogeneous settings. This leads us to the presentation of our method for online learning of the lattices in Subsection~\ref{ssec:olala}, which is followed by a discussion in Subsection~\ref{ssec:discussion}.

\subsection{\ac{fl} with Adaptive Lattice Quantizers}\label{ssec:FL_Adaptive}
We begin by describing a generic \ac{fl} setup where the users employ adaptive lattice-based compression. In this configuration, each communication round consists of both model updates and lattice quantizer metadata that evolve over time. 
Accordingly, the global model $\vw_t$ in \eqref{eq:FedAvg_update_rule}, which is the desired \ac{fa} outcome, and the local updates $\vh_t^u$, are respectively replaced with $\tilde\vw_t$ and $\tilde{\vh}_t^u$.
The procedure, detailed in Algorithm~\ref{alg:FL}, operates iteratively: in each round $t$, the  server broadcasts the current global model $\tilde\vw_{t-1}$ to all participating users; followed by user-side encoding and server-side decoding.

\subsubsection{User-Side Encoding}
Each user $u$ locally updates the model to obtain $\tilde\vh_t^u$ via, e.g., local \ac{sgd} \eqref{eq:local_sgd}. To reduce the cost of uplink communication, the update is compressed using a dithered lattice quantizer. The lattice chosen by user $u$ at round $t$, characterized by a generator matrix $\vG_t^u$ and support radius $\gamma_t^u$, is selected such that the resulting quantizer operates at a fixed target rate $R$. 
While here we do not specify how the choice of the lattice (i.e., $\vG_t^u$ and $\gamma_t^u$ in Step~\ref{stp:Adaptive} of Algorithm~\ref{alg:FL}) is done, we propose an online learning method for that aim in Subsection~\ref{ssec:olala}. The dither signal $\vd_t^u$ is independently generated using a shared pseudo-random seed $\xi_t^u$ (which can be set once when initializing the \ac{fl} procedure), enabling the server to reconstruct the same dither. 

Finally, the user transmits to the server the quantized model update $Q_{\mathcal{L}_{\gamma_t^u}(\vG_t^u)}\big(\tilde\vh^u_t + \vd_t^u\big)$ along with the metadata $\vG_t^u$. 
Note that, if we assume the metadata is sent with high-resolution, e.g., $64$ bits representation, the overall number of bits conveyed is $m \cdot R + 64\cdot L^2$, which is a minor overhead when training large models (i.e., large $m$).

\begin{algorithm}
    \caption{\ac{fl} with Adaptive Lattice Quantizers}
    \label{alg:FL}
    \SetKwInOut{Initialization}{Init}
    \Initialization{Data sets $\{\cD_u\}$, lattice dimension $L$,  rate $R$; \newline
    Initial model parameters $\vw_0$,  random seeds $\{\xi_t^u\}$}
    {
        \For{$t=1,2 \ldots$}{%
                    \nonl\textbf{Server side:}\\
                    Send global model $\tilde\vw_{t-1}$ to all users\; \label{stp:UserSel}

                    \For{$u =1,2,\ldots, U$}{
                    \nonl\textbf{User $u$ side:}\\
                        Obtain $\tilde\vh^u_{t}$  via local training on $\cD_u$ from $\tilde\vw_{t-1}$;  \label{stp:local}
                        
                        Set lattice $\vG_{t}^{u}$ and  $\gamma_t^u$ to quantize $\tilde\vh^u_{t}$  at rate $R$\; \label{stp:Adaptive}
    
                        Randomize dither $\vd_t^u$ with local seed $\xi_t^u$\; 
    
                        Send  $\vG_{t}^{u}$  and  $Q_{\cL_{\gamma_t^u}(\vG_{t}^{u})}\big(\tilde\vh^u_{t} + \vd_t^u\big)$\; 
                                      
                    \nonl\textbf{Server side:}\\
                        Compute $Q^{\rm SDQ}_{{\cL_{\gamma^t_u}(\vG^t_u)}}\big(\tilde\vh^u_{t}\big)$ via $\xi_t^u, \vd_t^u$\;

                    }
                    Aggregate model updates using~\eqref{eq:compressed_FedAvg_update_rule} to form $\tilde\vw_t$.                    
                    
                    }
        \KwRet{$\tilde\vw_t$}
  }
\end{algorithm}

\subsubsection{Server-Side Decoding}
Upon reception, the server reconstructs the dither vector $\vd_t^u$ using $\xi_t^u$ and recovers the compressed model via subtracting the dither, obtaining the \ac{sdq} of $\vw_t^u$. The global model is then updated by aggregating these quantized updates across all users, typically via \ac{fa}, i.e.,
\begin{equation}\label{eq:compressed_FedAvg_update_rule}
\tilde\vw_{t}=
\tilde\vw_{t-1} +  \frac{1}{U} \sum_{u=1}^{U} Q^{\rm SDQ}_{{\cL_{\gamma^t_u}(\vG^t_u)}}\left(\tilde\vh^u_{t}\right).     
\end{equation}
This routine allows each client to adapt its lattice across \ac{fl} rounds, enabling the quantizer to better match the current local update. Notably, due to the compact representation of the lattice via its generator matrix, the communication overhead associated with conveying the quantization metadata remains minimal.


\subsection{Theoretical Analysis}\label{subsec:analysis}
The adaptive scheme in Algorithm~\ref{alg:FL}, is next  analytically analyzed, revealing improved convergence behavior in heterogeneous \ac{fl} environments relative to its fixed, static counterpart.

\subsubsection{Updates Distortion}\label{subsubsec:weights_distortion}
Recall that the local updates transmitted from the users to the server in Algorithm~\ref{alg:FL} are  compressed via  lattice-based \ac{sdq}, where each edge device $u$ is associated with a different time-varying lattice generating matrix, $\vG^t_u$.
This quantization inherently induces some distortion, being introduced in the \ac{fl} training process. 
More formally, our goal is thus to quantify the difference between the `vanilla'-version $\vw_{t+1}$ in \eqref{eq:FedAvg_update_rule} to its dynamically-compressed alternative $\tilde\vw_{t+1}$ in~\eqref{eq:compressed_FedAvg_update_rule}.  

We next show that, under common assumptions used in \ac{fl} analysis, the effect of the excessive distortion induced by incorporating adaptive lattice quantizers can be mitigated while recovering the desired $\vw_{t+1}$ as $\tilde\vw_{t+1}$. Thus, the accuracy of the global learned model can be maintained. For ease of tractability, we focus on a the simple form of \lsgd \eqref{eq:local_sgd} with \ac{fa}, from which convergence guarantees of other forms of local updates (e.g.. mini-batch \lsgd, multiple local iterations) are known to follow~\cite{li2019convergence}. Accordingly,  the model updates in the following analysis are the stochastic gradients, namely, 
\begin{equation*}
\tilde{\vh}_t^u = -\eta_t \nabla F_u(\tilde{\vw}_t, \sample).    
\end{equation*}


To begin, we adopt the following assumptions on the local datasets and on the stochastic gradient vector $\nabla F_u(\cdot, \sample)$:
\begin{enumerate}[label={\em AS\arabic*},series=assumptions]
    \item \label{itm:heterogenity}
    Each dataset $\cD_u$ is comprised of i.i.d samples. However, different datasets can be statistically heterogeneous, i.e., arise from different distributions. 
   \item \label{itm:bounded_variance}
    The variance of stochastic gradients is bounded:
    \[\E \left[ \left\| \nabla F_u(\cdot, \sample) - \nabla F_u(\cdot) \right\|^2 \right] \leq \sigma_u^2, \quad u = 1, \ldots, U.\]
     \item \label{itm:zero_overeloading}
    The quantizer employed by each user is not overloaded, for each user $u$ and round $t$. 
\end{enumerate}
The statistical heterogeneity in \ref{itm:heterogenity}  is a common characteristic of \ac{fl} \cite{kairouz2021advances,shlezinger2020communication}. 
It implies that the  loss surfaces can differ between users, hence the dependence on $u$ in  \ref{itm:bounded_variance}, which is
often employed in distributed learning studies \cite{shlezinger2020uveqfed,li2019convergence, stich2018local, zhang2012communication}.

We can now bound the distance between the recovered (dynamically-quantized and stochastic) update  of \eqref{eq:compressed_FedAvg_update_rule} and the desired 'full' non-compressed and non-stochastic one $\frac{1}{U}\sum_{u=1}^U 
  \nabla F_u(\Tilde\vw_t)$, as stated in the following theorem:
\begin{theorem}[Distortion bound]\label{thm:bounded_var}   
When \ref{itm:heterogenity}-\ref{itm:zero_overeloading} hold, the expected distortion induced by \lsgd with lattice \ac{sdq} obeys
\begin{align}\label{eq:bounded_var}
&\E\left[\left\| 
\frac{1}{U}\sum_{u=1}^U 
 Q^{\rm SDQ}_{\cL_{\gamma_u^t}(\vG^t_u)}
 \left(\nabla F_u(\Tilde\vw_t, \sample)\right)
- 
\frac{1}{U}\sum_{u=1}^U 
\nabla F_u(\Tilde\vw_t)
\right\|^2 \right] \notag\\ 
& \qquad\qquad\leq
\frac{1}{U^2}\sum_{u=1}^{U}
\left(
\sigma^2_u + 
\sigma^2_{\rm SDQ}\left(\cL_{\gamma^t_u}(\vG^t_u)\right)
\right),
\end{align}
where $\sigma^2_{\rm SDQ}\left(\cdot\right)$ is given in~\eqref{eq:sigma2_sdq}. 
\end{theorem}
\begin{IEEEproof}
    The proof is given in Appendix \ref{app:bounded_var_proof}. 
\end{IEEEproof}
\smallskip    
Theorem \ref{thm:bounded_var} implicitly suggests that the recovered model can be made arbitrarily close to the desired one by increasing the number of edge users participating in the \ac{fl} training procedure. 
This is because we get that \eqref{eq:bounded_var} decreases as $1/U^2$.  
Besides, when the step-size $\eta_t$ gradually decreases, which is known to contribute to the convergence of \ac{fl} \cite{stich2018local, li2019convergence}, it follows from Theorem~\ref{thm:bounded_var} that the distortion decreases accordingly, further revealing its effect as the \ac{fl} iterations progress, discussed next. 

\subsubsection{Federated Learning Convergence}\label{subsubsec:fl_convergence}
To study the convergence of \ac{fa} with adaptive lattice quantizers, we further introduce the following assumption, inspired by \ac{fl} convergence studies in, e.g., \cite{shlezinger2020uveqfed,li2019convergence, stich2018local,lang2023joint, lang2023compressed, lang2024stragglers}:
\begin{enumerate}[resume*=assumptions]
    \item \label{itm:obj_smooth_convex}
    The local objective functions $\left\{F_u(\cdot)\right\}^U_{u=1}$ are all $L$-smooth and $\mu$-strongly convex, i.e., for all $\vw_1, \vw_2 \in \R^m$
    \begin{multline*}
        (\vw_1 -\vw_2)^T\nabla F_u(\vw_2)
        +\frac{1}{2}\rho_c{\|\vw_1 -\vw_2\|}^2 \\
        \leq  F_u(\vw_1)-F_u(\vw_2) \leq \\
        (\vw_1 -\vw_2)^T\nabla F_u(\vw_2)
        +\frac{1}{2}\rho_s{\|\vw_1 -\vw_2\|}^2.
    \end{multline*}
\end{enumerate}
Assumption \ref{itm:obj_smooth_convex} holds for a range of objective functions used in \ac{fl}, including $\ell_2$-norm regularized linear regression and logistic regression \cite{shlezinger2020uveqfed}. To proceed, following \ref{itm:heterogenity} and as in \cite{li2019convergence, shlezinger2020uveqfed,lang2023joint, lang2023compressed}, we define the heterogeneity gap,
\begin{align}\label{eq:psi_heterogeneity_gap}
\Gamma \triangleq F(\vw_{\rm opt})-\sum_{u=1}^U\alpha_u \min_{\vw} F_u(\vw),
\end{align}
where $\vw_{\rm opt}$ is defined in \eqref{eq:optimal_fl_model}. Note that \eqref{eq:psi_heterogeneity_gap} captures the degree of heterogeneity: if $\{\cD_u\}$ originate from the same distribution, $\Gamma$ tends to zero as the training size grows, and is positive otherwise.

The following theorem characterizes the convergence of  \ac{fl} with local \ac{sgd} training  using adaptive lattice quantizers:
\begin{theorem}[\ac{fl} convergence bound]\label{thm:FL Convergence}
Let Assumptions \ref{itm:heterogenity}-\ref{itm:obj_smooth_convex} hold and define $\kappa = \frac{L}{\mu}$, $\nu = \max\{8\kappa, 1\}$, and the learning rate 
$\eta_t = 2/\left(\mu(\nu + t)\right)$.
Then, for \ac{fl} with  users transmitting compressed gradients, that are quantized via lattice-based \ac{sdq} with a unique generating matrix per client, it holds that 
\begin{align}\label{eq:FL_Convergence}
 &   \E[F(\tilde\vw_T)] - F(\vw_{\rm opt}) 
    \notag \\
    & \leq \frac{\kappa}{\nu \!+ \!T\! - \! 1}\left(
    \frac{2}{\mu}\textcolor{black}{\max_{t=1,\dots,T}B_{t}} \!+\! \frac{\mu\nu}{2}\E\left[\|\tilde\vw_1 \!-\! \vw_{\rm opt}\|^2\right]
    \right),
\end{align}
where $\vw_{\rm opt}$ is defined in \eqref{eq:optimal_fl_model}, and 
\begin{equation}\label{eq:B_term_convergence}
B_{t} := \frac{1}{U^2}\sum_{u=1}^{U}
\left(\sigma^2_u + 
\sigma^2_{\rm SDQ}\left(\cL_{\gamma^t_u}(\vG^t_u)\right) 
\right)
+ 2L\Gamma.
\end{equation}
with  $\Gamma$ and  $\sigma^2_{\rm SDQ}\left(\cdot\right)$  defined in , \eqref{eq:psi_heterogeneity_gap} and \eqref{eq:sigma2_sdq}, respectively.
\end{theorem}
\begin{IEEEproof}
    The proof is given in Appendix \ref{app:FL Convergence_proof}. 
\end{IEEEproof}
\smallskip

Theorem \ref{thm:FL Convergence} rigorously bounds the difference in the objective value of the model learned via local \ac{sgd}-based \ac{fl} with adaptive lattice quantizers and the optimal model $\vw_{\rm opt}$, for any finite iteration index $T$. As such, it also frames the asymptotic convergence profile, i.e., the behavior of the bound in \eqref{eq:FL_Convergence} for arbitrarily large $T$.
Specifically, Theorem \ref{thm:FL Convergence} reveals that, when the sequence $\{B_t\}$ is bounded (and thus the distortion term is bounded), a convergence at a rate of $\cO(1/T)$ is achieved.
This asymptotic rate implies that as the number of iterations $T$ progresses, the  learned  model converges to $\vw_{\rm opt}$ with a difference decaying at the same order of convergence as \ac{fl} with no compression constraints \cite{stich2018local, li2019convergence}. 
Theorem \ref{thm:FL Convergence} thus indicates that adaptive lattice quantizers allows to incorporate compression considerations into non-i.i.d \ac{fl} to be done in a manner which does not increase the order of the asymptotic convergence profile.

Theorem~\ref{thm:FL Convergence} can be viewed as a generalization of \cite[Thm. 3]{shlezinger2020uveqfed}, which characterizes the convergence profile of \ac{sdq}-based \ac{fl} with a single fixed generating matrix for all the users. This generator matrix can either be arbitrary as in \cite{shlezinger2020uveqfed}, or optimized via \cite{lang2024data} or \cite{agrell1998optimization}. 
Therefore, if lattice-based compressed \ac{fl} can be shown to converge using a single fixed matrix across users and rounds, it raises the question of whether adaptively learning it is beneficial. This question  is addressed next.



\subsubsection{Convergence Bound Minimization}
While the convergence rate is of asymptotic order $\cO(1/T)$, the need to compress the model updates affects the convergence of the model in the non-asymptotic regime. This is revealed in Theorem~\ref{thm:FL Convergence} via the coefficient $B_t$ in \eqref{eq:FL_Convergence}. This coefficient includes several additive terms, with each stemming from a different consideration affecting the learning profile. For instance, the term depending on $\Gamma$ indicates that statistical heterogeneity makes convergence slower, as noted in \cite{li2019convergence}. This heterogeneity is further captured in \eqref{eq:B_term_convergence} via the averaged \ac{sdq} \ac{mse} distortion, i.e., $\sum_{u=1}^{U}
\sigma^2_{\rm SDQ}\left(\cL_{\gamma^t_u}(\vG^t_u)\right)$. 
In this context, convergence is influenced by all lattices up to time step~$T$, with the set incurring the highest error having the dominant effect.
Consequently, adaptive lattices are beneficial if by adopting non-identical lattices we result with a lower convergence bound compared to the one that is obtained with identical lattices. Formally, we wish to show that $\forall t,u$ for every fixed  $\vG$, there exists a sequence of lattices $\{\vG_u^t\}$ such  $\sigma^2_{\rm SDQ}\left(\cL_{\gamma^t_u}(\vG^t_u)\right)\leq \sigma^2_{\rm SDQ}\left(\cL_{\gamma^t_u}(\vG)\right)$, implying that the convergence bound is minimized by allowing the lattices to vary. This is rigorously stated in the following theorem:

\begin{theorem}\label{thm:gamma_dependent_optimal_G}
Consider the quantizer of Definition~\ref{def:lattice_quantizer}, using a $\gamma$-radius truncated lattice with generating matrix $\vG$. Now, assume that the quantizer is not overloaded, and that the quantization rate is $R$. Then, the generating matrix which minimizes the \ac{mse} distortion of the \ac{sdq} scheme depends on $\gamma$ , i.e.,
\begin{equation}\label{eq:gamma_dependent_optimal_G}
   \argmin_{\vG} 
\sigma^2_{\rm SDQ}(\cL_\gamma(\vG))
=
\gamma^2\cdot \argmin_{\vA \in \mathrm{SL}_L(\mathbb{R})} \cF(\vA,R),
\end{equation}
where $\mathrm{SL}_L(\mathbb{R})$ is the set of real $L \times L$ matrices with  $\det(\cdot) = 1$; and $\cF(\vA,R)$ denotes a function depending solely on $\vA$ and $R$.
\end{theorem}
\begin{IEEEproof}
    The proof is given in Appendix \ref{app:gamma_dependent_optimal_G_proof}. 
\end{IEEEproof}
\smallskip
To better grasp the role of $\gamma$, recall that the data is assumed to be heterogeneous among the users, and so is the distribution of the induced model weights to be compressed before transmitted to the \ac{fl} server. Moreover, as the model evolves over time, this distribution is also expected to vary across time steps. Therefore, $\gamma$ depends on both \ac{fl} users and rounds.
As this work considers compressing the updates via \ac{sdq}, $\gamma$ can be viewed as an adaptive dynamic range, chosen such that it guarantees zero-overloading, or equivalently a scaling coefficient applied to the model updates; assuring the validity of Theorem~\ref{thm:SDQ}. In contrast, the minimizer of $\cF(\vA, R)$ is purely relied on geometrical aspects, and can therefore be identical for all the users. In particular, its asymptotic regime minimizer, where $R\to \infty$, is studied in classical information theory literature, e.g., \cite{zamir1996lattice}.

In a broader perspective, Theorem~\ref{thm:gamma_dependent_optimal_G} implies that for  heterogeneous \ac{fl}, it is better (in terms of model convergence) to deviate from using identical quantization rules. This deviation, though, is not reflected in, e.g., the design of the quantizer decision cells, but rather in forming the generating matrix in a user-specific per-round manner. This is captures by \eqref{eq:gamma_dependent_optimal_G}, which reflects the interplay between the two core aspects associated with the scheme of \ac{sdq}: input-distribution invariance (i.e., universality) and zero-overloading assumption (Theorem~\ref{thm:SDQ}). 
Theorem~\ref{thm:gamma_dependent_optimal_G}  also indicates on the potential of  allowing certain amount of overloading, as preventing it directly quadratically scales the associated error. This insight and its numerical consequences on the \ac{fl} model performance and convergence are systemically showcased in Section~\ref{sec:experiments}.


\subsection{\ac{name} Algorithm}
\label{ssec:olala}
The theoretical analysis reported in the previous subsection does not consider  how the local lattices  $\{\vG_t^u\}$ are chosen in Step~\ref{stp:Adaptive} of Algorithm~\ref{alg:FL}. Still, its findings reveal two key considerations as to the importance of this step in the \ac{fl} procedure:
\begin{enumerate}[label={\em C\arabic*}]
    \item  \label{itm:NeedAdaptive} The fact that in \ac{fl} (and particularly in heterogeneous settings), the distribution of the  local updates can vary  between rounds ($t$) and users ($u$) indicates that one can enhance the learning procedure by properly altering $\{\vG_t^u\}$ over $t$ and $u$.
    \item \label{itm:NeedLearn} The theoretical analysis assumes that the lattices are not overloaded (via \ref{itm:zero_overeloading}), which allows to rigorously characterize the distortion via Theorem~\ref{thm:SDQ}. When allowing some level of overloading, one can potentially achieve less distorted quantization (particularly when operating with limited quantization rates) for which, as opposed to the non-overloaded case, there is no tractable  characterization. 
\end{enumerate}

\begin{figure*}
    \centering
    \includegraphics[width=1\textwidth]{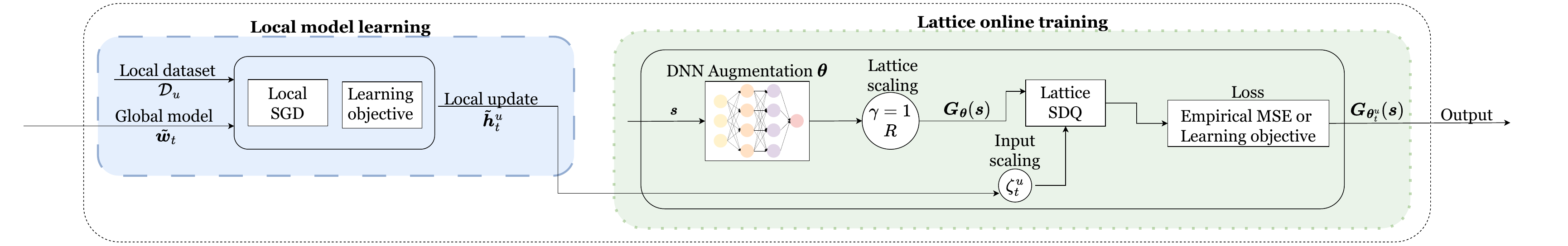}
    \caption{Overview of \ac{name} at user $u$ at time-step $t$, performing local model (left) and lattice (right) learning using input $\tilde{\vw}_t$ to form the  generator matrix $\vG_{\myVec{\theta}_t^u}(\vs)$. 
     }
    \label{fig:architecture}
\end{figure*}

The identification of \ref{itm:NeedAdaptive} motivates us to develop a scheme to implement Step~\ref{stp:Adaptive} of Algorithm~\ref{alg:FL} via online lattice adaptation in a manner that can be carried out independently by each user on each round. Based on \ref{itm:NeedLearn}, we opt a machine learning approach for optimizing the lattices while allowing some level of overloading. To describe the resulting {\em \ac{name}} algorithm, illustrated in Fig.~\ref{fig:architecture}, we first discuss how we cast the local lattices as the trainable parameters of a machine learning architecture, after which we explain the loss that guides their tuning and the online learning procedure. 

\subsubsection{Architecture}
We leverage deep learning tools for tuning the lattice generator matrix of user $u$ at round $t$, denoted $\vG_t^u$. To that aim,  we formulate the lattice quantizer as a {\em trainable machine learning model}, parameterized by $\myVec{\theta}_t^u$. This formulation is comprised of \ac{dnn} augmentation and an alternative representation of the quantization operation.

{\bf \ac{dnn} Augmentation:} Instead of treating $\vG_t^u$ as a trainable parameter, we follow the deep prior approach of \cite{ulyanov2018deep}, and set it to be the output of a \ac{dnn} with weights $\myVec{\theta}_t^u$, {\em fixed} input $\vs$, and $L^2$ output neurons. The reshaped \ac{dnn} output is scaled to have at most $2^{LR}$ codewords within radius $\gamma$, and the resulting matrix, denoted $\vG_{\myVec{\theta}_t^u}(\vs)$, is used as the  generator matrix. This augmentation exploits the abstractness of \acp{dnn}, while stabilizing and facilitating the learning procedure.

For simplicity and following \cite{shlezinger2020uveqfed}, we fix the quantization radius to $\gamma =1$, and replace its role with a scaling parameter, denoted $\zeta_t^u$. The model updates are scaled by $\zeta_t^u$ to achieve a desired level of overloading (less that $1\%$) within $\gamma$. This necessitates conveying  $\zeta_t^u$ to the server, in addition to the \ac{dnn} output $\vG_{\myVec{\theta}_t^u}(\vs)$ and the quantized updates.


{\bf Reformulated Lattice Quantization:} The formulation of \eqref{eqn:latticeQuant} limits the ability to apply gradient-based optimization due to its continuous-to-discrete nature. Common techniques in the deep learning literature for dealing with quantized features, e.g., straight-through estimation~\cite{huh2023straightening}, are designed to facilitate backpropagation through non-differentiable mappings (e.g., obtain the gradient of the loss with respect to the quantizer input), while we are interested in optimizing parameters of the quantizer itself. For this reason, previous studies on learned quantization~\cite{shlezinger2022deep,danial2025learning}, including our preliminary work~\cite{lang2024data}, utilized a {\em differentiable approximation} of the quantizer. These approaches require careful tuning of additional hyperparameters to achieve reliable learning, which can be done in offline training (where one can explore various hyperparameters), but is not suitable for our online learning setting. 

To cope with this, we introduce an {\em exact} reformulation of \eqref{eqn:latticeQuant} which supports gradient-based learning. To that aim,  for a  (possibly finite) lattice with generator matrix $\vG$,  denoted $\cL(\vG)$, we define the set of lattice point indices 
\begin{equation}
\label{eqn:DefIdxSet}
 \cI_{\cL(\vG)} \triangleq \{\vl  \in \mathbb{Z}^L : \vG \vl \in \cL(\vG)\}.
\end{equation}
Using \eqref{eqn:DefIdxSet}, we  equivalently write the lattice quantizer  \eqref{eqn:latticeQuant} as 
\begin{equation}
\label{eqn:latticeQuantEq}
    Q_{\cL(\vG)}(\vx) = \vG \argmin_{\vl \in  \cI_{\cL(\vG)} }\|\vx-\vG \vl\|. 
\end{equation} 
The reformulation of \eqref{eqn:latticeQuant} as \eqref{eqn:latticeQuantEq} still includes the non-differentiable $\argmin$ operator. However, it allows one to form a computation graph that enables training the parameters dictating $\vG$ via gradient-based methods due to its {\em linear dependence} of $\vG$. Specifically, using \eqref{eqn:latticeQuantEq}, one can approximate 
\begin{equation}
\label{eqn:approximated_grad}
   \nabla_{\myVec{\theta}} Q_{\cL(\vG_\myVec{\theta}(\vs))}(\vx) \approx \left(\nabla_{\myVec{\theta}} \vG_\myVec{\theta}(\vs)\right) \argmin_{\vl \in  \cI_{\cL(\vG_\myVec{\theta}(\vs))} }\|\vx-\vG_\myVec{\theta}(\vs)\vl\|,
\end{equation} 
in  which the $\arg \min$ statement can be viewed as a form of {\em stop gradient} \cite{van2017neural}.

\subsubsection{Loss Function}
Since \ac{name} is designed to adapt the local lattices by casting their operation as a machine learning architecture, their tuning requires  formulating a data-driven loss function. Note that  the setting of the local lattices is carried out in Step~\ref{stp:Adaptive} of Algorithm~\ref{alg:FL}, i.e., after local training (Step~\ref{stp:local}). Thus, the data used for online learning by user $u$ at round $t$ is the current global model $\tilde\vw_{t}$ and its local update  $\tilde\vh^u_{t}$. 

Using the  model and its  local updates as online learning data, we formulate two candidate empirical risk objectives:
\begin{itemize}
    \item {\em Empirical \ac{mse}}, i.e., minimize quantization distortion, 
    \begin{equation}
    \label{eqn:lossMSE}
       \tilde{F}_{\vw, \vh }^u(\myVec{\theta}) =  \left\| {\vh} -  Q^{\rm SDQ}_{\cL_{\gamma}(\vG_\myVec{\theta}(\vs))}({\vh})   \right\|^2.
    \end{equation}
    \item {\em Learning Objective} using the local \ac{fl} losses (as in \eqref{eq:optimal_fl_model})  
    \begin{equation}
    \label{eqn:losslearn}
       \tilde{F}_{\vw}^u(\myVec{\theta}) = F_u\left(\vw+ Q^{\rm SDQ}_{\cL_{\gamma}(\vG_\myVec{\theta}(\vs))}(\vh) \right)  .
    \end{equation}
\end{itemize}
The loss in \eqref{eqn:losslearn} boosts a form of {\em task-based quantization}, as the quantizer is tuned based on the overall learning task, as opposed to the \ac{mse} loss in \eqref{eqn:lossMSE}, which seeks standard distortion minimization. However, as we view the local model as our data for online learning, \eqref{eqn:lossMSE} has the gain of supporting mini-batch based optimization, as it can be computed for any set of sub-vectors of $\tilde\vw^u_{t}$, while \eqref{eqn:losslearn} requires  the entire vector $\tilde\vw^u_{t}$. 

\subsubsection{Online Training}
The loss functions  enable unsupervised online learning of the lattices, as, e.g., no ground-truth digital representation is required. The ability to compute the gradients with respect to the lattice generator matrix via \eqref{eqn:approximated_grad} enables training using conventional deep learning optimizers based on first-order methods. An example using mini--batch \ac{sgd} (which, when using the loss \eqref{eqn:losslearn}, is applicable with $\beta=1$ batch per epoch) is formulated as Algorithm~\ref{alg:training}. 
Once training is concluded, i.e., $\myVec{\theta}_t^u$ is learned, the resulting $\vG_{\myVec{\theta}_t^u}(\vs)$ is used for lattice \ac{sdq} in Step~\ref{stp:Adaptive} of the overall Algorithm~\ref{alg:FL}.

\begin{algorithm}
\caption{Online Lattice Learning at User $u$, Round $t$}
\label{alg:training} 
\SetKwInOut{Initialization}{Init}
\Initialization{Randomize $\myVec{\theta}$, fix learning rate $\eta$, epochs $i_{\max}$, input $\vs$, and number of batches $\beta$}
\SetKwInOut{Input}{Input}  
\Input{Current model $\tilde\vw_{t}$ and update  $\tilde\vh^u_{t}$ }  
{\For{$i = 0 \ldots, i_{\max}-1$}{%
                Randomly divide input into $L$-length $M$ sub-vectors\;
                Randomly divide $M$-size data into batches $\{\vw_b, \vh_b\}_{b=1}^\beta$\;
                \For{$b =0, \ldots, \beta-1$}{
                                  
                    Compute batch loss $ \tilde{F}_{\vw_b, \vh_b}^u(\myVec{\theta})$ via \eqref{eqn:lossMSE} or \eqref{eqn:losslearn}\;
                    Update  $\myVec{\theta}\leftarrow \myVec{\theta} - \eta\nabla_{\myVec{\theta}}\tilde{F}_{\vw_b, \vh_b}^u(\myVec{\theta})$\;
                    }          
                }                
                }
                Set $\myVec{\theta}_t^u\leftarrow \myVec{\theta}$\;
\SetKwInOut{Output}{Output}                  
    \Output{Generator matrix $\vG_{\myVec{\theta}_t^u}(\vs)$}
\end{algorithm}

\subsection{Discussion}
\label{ssec:discussion}
The proposed \ac{name} framework introduces a novel mechanism for incorporating adaptive lattice quantization in \ac{fl}. Our key insight is the identification of the value in tailoring the lattice quantizer to the local update  at each user and communication round, thereby enabling more accurate and efficient model update transmission. \ac{name} enables such adaptation by allowing each user to learn and apply a custom lattice, represented compactly through its $L \times L$ generator matrix. This compact representation is independent of the quantization rate (as opposed to generic vector quantizers, whose representation requires sharing a complete rate-specific codebook), rendering the communication overhead of transmitting the quantizer metadata negligible, particularly in the context of training large-scale \acp{dnn}. Beyond its practical advantages, the use of structured lattice codes also introduces theoretical tractability into the analysis of the resulting quantization error and its propagation through the \ac{fl} training process.

From a computational standpoint, the added complexity introduced by \ac{name} over fixed-lattice \ac{fl}  stems primarily from the need to learn the generator matrix during local training. This involves evaluating a compact neural network and computing its gradients on limited data (the model weights), which is modest in scale compared to the training of deep models on large data sets, as already done locally in \ac{fl}. As such, the overhead is minimal when assessed over typical \ac{fl} training routines, where the majority of computational effort remains focused on optimizing the global model parameters. Furthermore, our implementation ensures that the generator matrix can be efficiently updated using standard optimization routines with minor additional latency.

\ac{name} offers multiple modes of operation depending on the learning objective and desired granularity of adaptation. We investigate two training losses for learning the lattice: a learning-oriented loss that encourages \ac{fl} performance, and a distortion-oriented \ac{mse} loss that directly minimizes quantization error. Our empirical results indicate that both losses are effective, with the \ac{mse} loss often yielding better quantizers in terms of reconstruction error,  at the cost of increased sensitivity to optimization instability. Flexibility is also provided in the spatial and temporal granularity of lattice learning. \ac{name} supports per-user-per-round adaptation, but can also be restricted to learning a fixed lattice per user or even a globally shared lattice across all users, potentially via offline training. Nonetheless, our results consistently show that the dynamic, online adaptation offered by \ac{name}  leads to significantly improved performance over these static alternatives, justifying the added training complexity.

Several potential extensions of \ac{name}  are left for future work. In our implementation, the number of codewords in the lattice is controlled via a scaling radius $\gamma$, but one may consider alternative mechanisms to enforce a fixed codebook size, such as entropy-constrained quantization~\cite{chou1989entropy} or volume-preserving projections. Additionally, we learn the generator matrix indirectly via a compact neural network with a fixed input to ensure numerical stability and regularization. Future studies may explore direct optimization of the generator matrix, adopt alternative parameterizations that improve convergence or expressivity, and even consider designing machine learning models to map the weights into the lattice generator matrix. Finally, \ac{name} currently assumes a fixed distribution for the subtractive dither; extending the framework to learn or adapt the dither distribution in conjunction with the lattice may further enhance performance and robustness.

\section{Experimental Study}\label{sec:experiments}
We next numerically assess heterogeneous \ac{fl} with adaptive lattice earning via \ac{name}\footnote{The source code used in our experimental study, including all the hyper-parameters, is available online at \url{https://github.com/Maya-Simhi/OLALa}.}. 
We first detail the experimental setup in Subsection~\ref{ssec:ExpSetup}; after which we evaluate the performance of the learned lattices in Subsection~\ref{ssec:Ablation}, as well as that of the \ac{fl} global model trained using them in Subsection~\ref{ssec:Results}.
\subsection{Experimental Setup}
\label{ssec:ExpSetup}

\subsubsection{Learning Tasks and Data Partitioning}

To evaluate the performance of OLALa, we consider two canonical \ac{fl} tasks with distinct data sets and architectures: handwritten digit classification using MNIST and natural image classification using CIFAR-10.

{\bf MNIST Classification}:
The first task involves federated training over the MNIST dataset, which consists of $28 \times 28$ grayscale images of handwritten digits. The dataset includes 60,000 training samples and 10,000 test samples. Local training on each user is carried out using \ac{sgd} with a learning rate of $0.1$, and the classification loss is computed using the standard cross-entropy loss. We examine three different \acp{dnn}:
\begin{itemize}
    \item {\em Linear}: A simple linear regression model.
    \item {\em MLP}: A fully-connected \ac{dnn} with two hidden layers and ReLU activations.
    \item {\em CNN}: A \ac{dnn} with two convolutional layers followed by two fully connected layers with ReLU activations and max-pooling.
\end{itemize}
The \ac{fl} process is run for 40 global communication rounds, each comprising 100 local \ac{sgd} steps. Lattice adaptation in \ac{name} is performed every 10 local steps.

{\bf CIFAR-10 Classification}:
The second task uses the CIFAR-10 dataset, which comprises $32 \times 32$ RGB images categorized into 10 natural image classes, with 50,000 training and 10,000 test examples. The trained model is a CNN consisting of three convolutional layers, followed by four fully connected layers, with ReLU activations, max-pooling, and dropout layers. Local training employs \ac{sgd} with a learning rate of $0.003$, using cross-entropy loss. The system is run for 100 global communication rounds, each containing 1500 local iterations. Lattice learning in \ac{name} is updated every 100 local steps.

\smallskip
To emulate user heterogeneity, both tasks employ a non-i.i.d. data partitioning across $U = 5$ users. Each user is assigned samples from three classes, with a one-class overlap between consecutive users. Specifically, user $u$ (indexed from 0 to 4) is assigned classes $\{2u, 2u+1, 2u+2\}$, modulo 10. This partitioning ensures both sufficient local variation and partial class overlap, achieving heterogeneous \ac{fl}  while ensuring that each user's local learning impacts the global model. The setup is particularly useful for evaluating the effectiveness and fairness of user-specific quantization, such that no individual client becomes disproportionately detrimental to the training performance due to mismatched quantization.

\subsubsection{Benchmarks}

To evaluate the effectiveness of OLALa, we compare it against several \ac{fl} baselines that represent different strategies for lattice quantization. For all lattice quantizers, we use dimension $d=2$, and apply the same scaling scheme as in \ac{name}, i.e., we scale the generator matrix to achieve the  quantization rate (which differs between tests) within a unit radius, and scale the input to meet overloading probability of up to $0.5\%$. 

{\bf Non-Compressed FL}:
As an upper-bound benchmark, we consider an \ac{fl} setup with no quantization. 
This represents the best achievable performance in terms of accuracy, but with maximal communication overhead.

{\bf Fixed Lattice Quantization}:
This baseline uses predefined lattice quantizers that remain fixed across all users and throughout the entire training process. We consider three generator matrices taken from the lattice quantization literature~\cite{nebe2025lattices}: 
\begin{itemize}
    \item {\em Hexagonal} lattice $\vG_{\rm Hex} =  \begin{bmatrix}
 1 & \frac{1}{2} \\
 0 & \frac{\sqrt{3}}{2}
 \end{bmatrix}$. 
 \item $D_2$ lattice $\vG_{D_2} = \begin{bmatrix}
 2 & 0 \\
 1 & -1
 \end{bmatrix}$.
 \item   $A_2$ lattice $\vG_{A_2} = \begin{bmatrix}
\sqrt{2} & 0 \\
 -0.7071 & 1.2247
 \end{bmatrix}$. 
\end{itemize}

{\bf Learned Static Lattices}:
This baseline includes two variants of \ac{name} in which the lattice generator matrix is learned once using data-driven optimization, but remains fixed throughout the \ac{fl} process. We consider two types of learned static lattices:
\begin{itemize}
    \item \emph{Global static}, where a single lattice is learned offline and used by all users. 
    \item \emph{User-dependent static}, where each user is assigned a fixed but individually learned lattice based on its local data.
\end{itemize} 
These benchmarks allow evaluating the individual gains in adaptivity across users and rounds of \ac{name}.






\subsection{Adaptive Lattice Learning Evaluation}
\label{ssec:Ablation}
\begin{figure}
    \centering
    \includegraphics[width=1\linewidth]{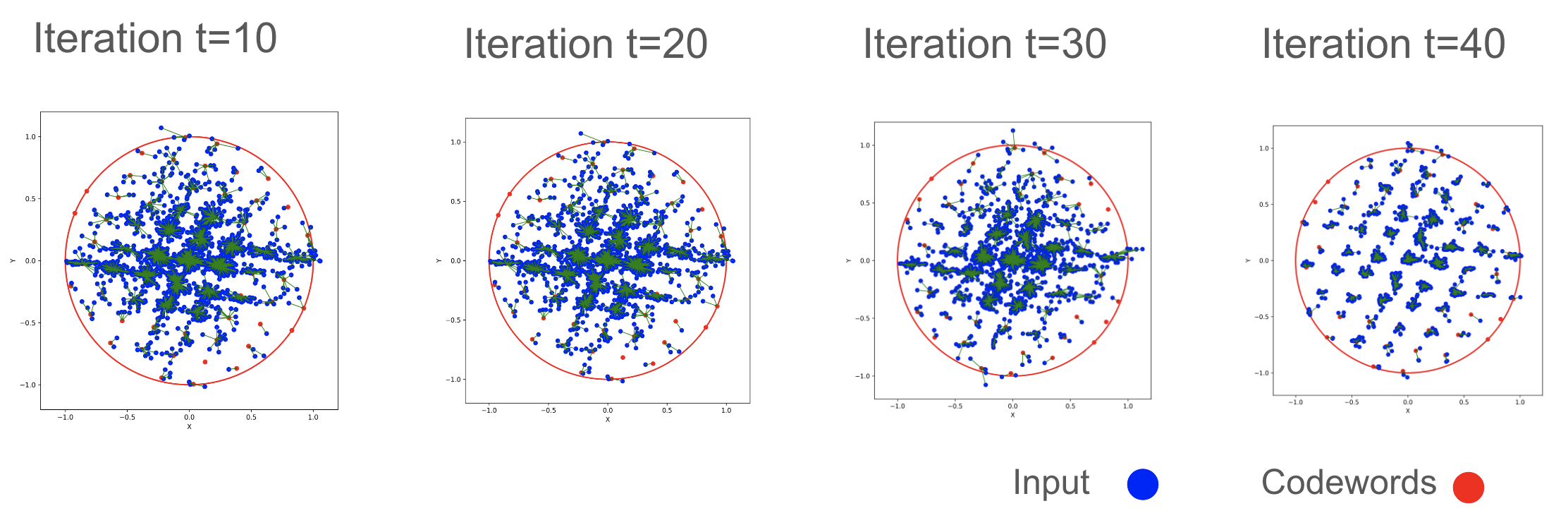}
    \caption{Evolution of OLALa's lattice quantizer over training rounds for User~1 on MNIST. Blue dots represent scaled weights, red dots denote codewords, red circle is the support radius $\gamma$, and green lines connect each weight to its nearest codeword.}
    \label{fig:learningMatrix}
\end{figure}

{\bf Lattice Adaptation}:
We begin our empirical study by investigating the internal behavior of \ac{name}, focusing on its ability to learn the quantization lattice in an online manner. 
To that aim, we first assess how OLALa adapts the lattice generator matrix over time, by visualizing the learned quantizer of a representative user during training on  MNIST. We use a quantization rate of $R=3$ (64 codewords at $d=2$) and plot the lattice configuration every 10 global communication rounds. 
As shown in Fig.~\ref{fig:learningMatrix}, each plot depicts the scaled local model weights (blue), the quantizer codewords (red), the support radius $\gamma$ (red circle), and the connection between each weight and its nearest codeword (green lines).

The results clearly illustrate that the generator matrix evolves throughout the learning process. Early in training, the codewords are not well aligned with the distribution of the model weights, leading to visible mismatches (green lines). As training progresses, the quantizer becomes increasingly aligned with the model weights: the codewords spread more effectively across the input space, and the number and length of the green lines decrease. These observations support the conclusion that our online adaptation  enables accurate and  refined quantization.

{\bf Adaptation Loss}:
We next evaluate how the choice of loss function used to guide lattice adaptation affects performance. Specifically, we compare three candidate loss measures: 
the learned objective (accuracy) as in \eqref{eqn:losslearn}; the \ac{mse} as in \eqref{eqn:lossMSE}; and a similar distortion-oriented loss, given by the (negative) \ac{snr}, defined as  
    $$ -{\rm SNR} = -\|{\vh}\|^2 / \left\| {\vh} -  Q^{\rm SDQ}_{\cL_{\gamma}(\vG_\myVec{\theta}(\vs))}({\vh})   \right\|^2.$$
While the accuracy-based loss is task-oriented, SNR and MSE directly target the distortion induced by quantization.

Table~\ref{tab:lrVsLoss} reports the resulting performance for different learning rates when training a CNN model on MNIST at $R=3$. All three loss functions yield useful gradients for optimizing the lattice, but their behaviors differ significantly. The objective-based loss leads to relatively stable performance across a broad range of learning rates, indicating robustness to hyperparameter settings. In contrast, distortion-oriented objectives (SNR and MSE) can achieve higher performance when properly tuned, but exhibit greater sensitivity to the learning rate.
These findings suggest that while distortion-based objectives may offer improved quantization quality, care must be taken when selecting the learning rate to avoid instability. Accuracy-based objectives, though less aggressive in minimizing distortion, provide a stable and reliable alternative.

\begin{table}
\centering
\begin{tabular}{lccc}
\toprule
Learning Rate & Accuracy & SNR & MSE \\
\midrule
$1\text{e}{-9}$ & 81.43\% & 71.60\% & 76.94\% \\
$1\text{e}{-8}$ & 81.43\% & 82.85\% & 84.01\% \\
$1\text{e}{-7}$ & 81.86\% & 89.61\% & 84.51\% \\
$1\text{e}{-6}$ & 81.43\% & 87.35\% & 87.08\% \\
$1\text{e}{-5}$ & 81.43\% & 85.21\% & 87.12\% \\
$1\text{e}{-4}$ & 81.43\% & 90.07\% & 84.67\% \\
$1\text{e}{-3}$ & 81.43\% & 84.98\% & 76.72\% \\
\bottomrule
\end{tabular}
\caption{Accuracy  after $40$ rounds with \ac{name} trained with different  loss functions, MNIST (CNN model, $R=3$). }
\label{tab:lrVsLoss}
\end{table}

{\bf Quantization Overloading}: 
We proceed to investigate the impact of allowing a controlled degree of  overloading on the final learning accuracy. While traditional lattice quantization designs aim to entirely avoid overloading to ensure bounded distortion, recent insights suggest that relaxing this constraint may provide performance benefits.
To explore this trade-off, we evaluated the classification accuracy of the CNN model trained on MNIST using \ac{name} and the various baseline quantization schemes, while varying the allowable overloading level. Each setting constrains the percentage of local model weights that are permitted to fall outside the quantizer support radius $\gamma$. In addition to fixed percentage-based thresholds (ranging from $0\%$ to $50\%$). 
We include a hybrid heuristic denoted by ``$-1$'', which limits overloading to at most $0.3\%$ of the update sub-vectors  not deviating from their empirical mean by more than $3\times$ their empirical standard deviation. 

The results, summarized in Table~\ref{tab:overloading}, reveal several important insights. First, strictly avoiding overloading ($0\%$) leads to noticeably degraded performance across all methods, due to excessive clipping or over-conservative scaling. Conversely, permitting large amounts of overloading (e.g., $10\%-50\%$) harms performance due to unbounded distortion. Moderate overloading levels (around $0.5\%-1\%$) strike a better balance, but the best overall performance is typically achieved with the heuristic ``$-1$'' approach, which combines principled statistical filtering with a tight bound on overload count. These results validate the design choice in \ac{name} to allow small, controlled overloading during quantizer adaptation.

\begin{table}
\centering
\scriptsize
\begin{tabular}{lcccccc}
\toprule
Method   & $-1$     & $0\%$     & $0.5\%$   & $1\%$     & $10\%$    & $50\%$    \\
\midrule
\ac{name}   & 93.01\%  & 87.05\%   & 92.06\%   & 90.38\%   & 86.28\%   & 65.70\%   \\
Static-Each     & 92.78\%  & 86.80\%   & 91.95\%   & 90.49\%   & 87.71\%   & 68.17\%   \\
Static-Global   & 91.59\%  & 87.78\%   & 90.91\%   & 91.73\%   & 90.06\%   & 69.22\%   \\
Fixed-Hexagon  & 90.36\%  & 86.53\%   & 88.78\%   & 91.43\%   & 92.89\%   & 71.65\%   \\
\bottomrule
\end{tabular}
\caption{Final test accuracy of different quantization methods on MNIST (CNN ,$R=3$) under varying overloading thresholds. ``$-1$'' denotes a heuristic allowing at most  $0.05\%$ overloading with minimum variance filtering.}
\label{tab:overloading}
\end{table}

\subsection{Federated Learning Results}
\label{ssec:Results}

We now evaluate the effectiveness of \ac{name} in full federated training, comparing its performance to that of alternative quantization schemes. Two key studies are reported: the first quantifies the final learning performance achieved with varying quantization rates; the second analyzes the training dynamics over time across different learning models and datasets.

{\bf Performance vs. Quantization Rate}:
In this study, we assess how the quantization rate used in the lattice quantizer affects the performance of the global model. We report both the classification accuracy (averaged over the last five global rounds) and the \ac{snr} of the quantized model updates. The results are presented in Table~\ref{tab:accVsCodewords} for the MNIST dataset using the CNN architecture.

Several trends emerge from Table~\ref{tab:accVsCodewords}. First, all quantized methods show improved performance as the number of codewords increases, approaching the baseline performance of the uncompressed model ($95.24\%$). Second, learning non-identical lattices, whether performed once per user (``Static-Each'') or adaptively during training (``\ac{name}''), outperforms both global and fixed lattice baselines. Notably, \ac{name} consistently achieves the best results across all quantization rates. Finally, we observe that a higher SNR does not always correlate with better accuracy; this discrepancy may stem from differences in how distortion is distributed across model layers and highlights the importance of task-aware quantizer optimization.

\begin{table}
\centering
\scriptsize
\begin{tabular}{lcccc}
\toprule
\textbf{Rate $R$} & $2$ & $2.5$ & $3$ & $3.5$ \\
\midrule
Non-comp.        & \multicolumn{4}{c}{95.24} \\
\midrule
\ac{name}       & 81.76 / 14.56 & 90.42 / 21.86 & 93.00 / 17.63 & 94.61 / 20.85 \\
Static-Each        & 80.41 / 12.23 & 87.51 / 14.53 & 92.78 / 16.33 & 94.40 / 18.54 \\
Static-Global      & 79.74 / 24.39 & 85.31 / 22.49 & 91.59 / 22.35 & 94.22 / 24.46 \\
Fixed-Hexagon     & 57.71 / 26.61 & 84.63 / 20.69 & 90.36 / 22.16 & 93.29 / 24.92 \\
Fixed-A2          & 74.25 / 20.29 & 85.96 / 21.09 & 90.78 / 22.78 & 93.43 / 24.29 \\
Fixed-D2          & 77.97 / 19.00 & 87.26 / 16.77 & 90.19 / 22.79 & 92.89 / 21.38 \\
\bottomrule
\end{tabular}
\caption{Final accuracy (in $\%$) / SNR for different quantization schemes on MNIST (CNN) with varying codeword counts. Accuracy is averaged over the final 5 rounds.}
\label{tab:accVsCodewords}
\end{table}

\begin{figure}
    \centering
    \includegraphics[width=\linewidth]{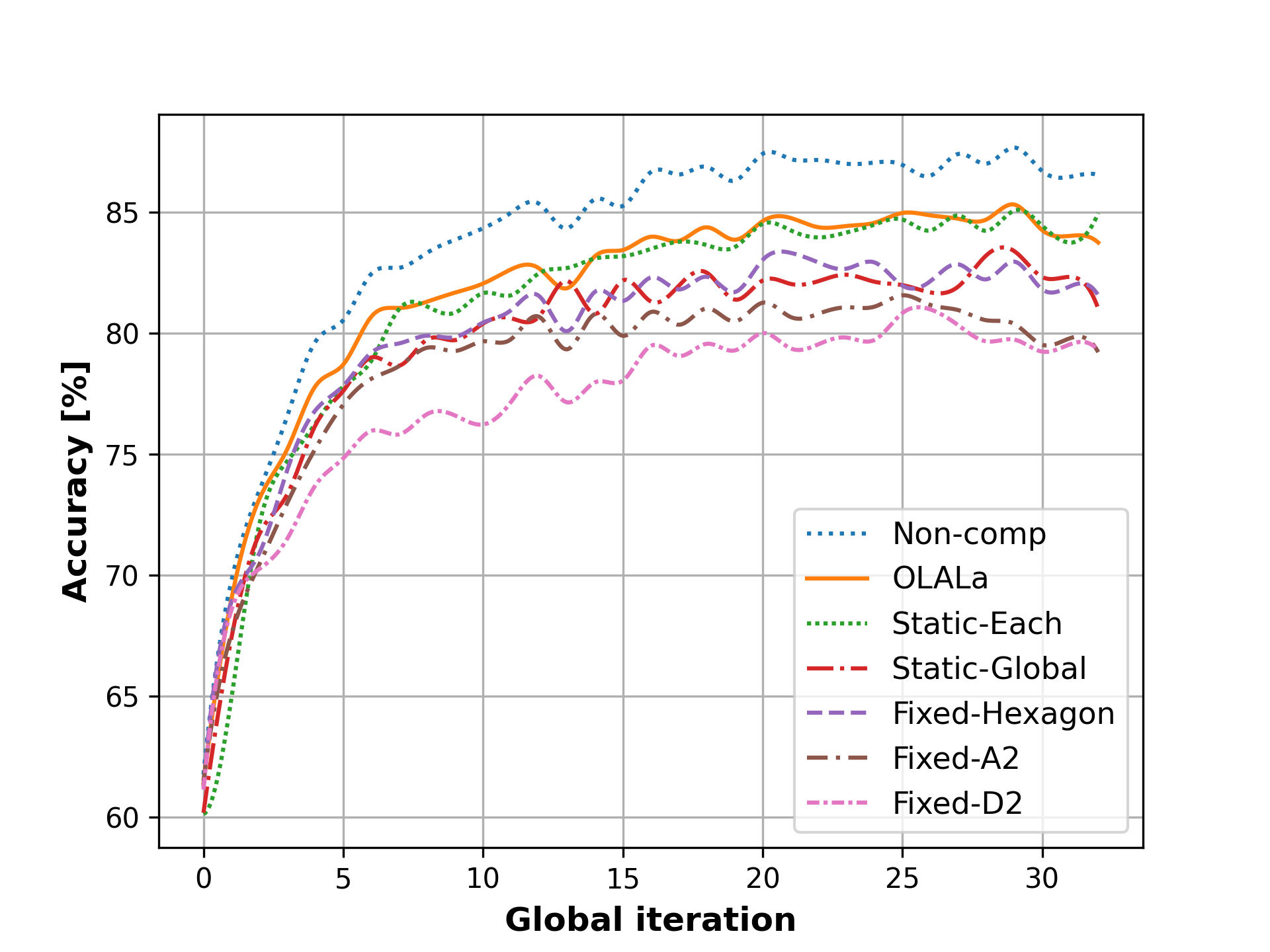}
    \caption{Accuracy vs. training rounds, MNIST,  linear model, $R=3.5$.}
    \label{fig:mnistLinear}
\end{figure}

\begin{figure}
    \centering
    \includegraphics[width=\linewidth]{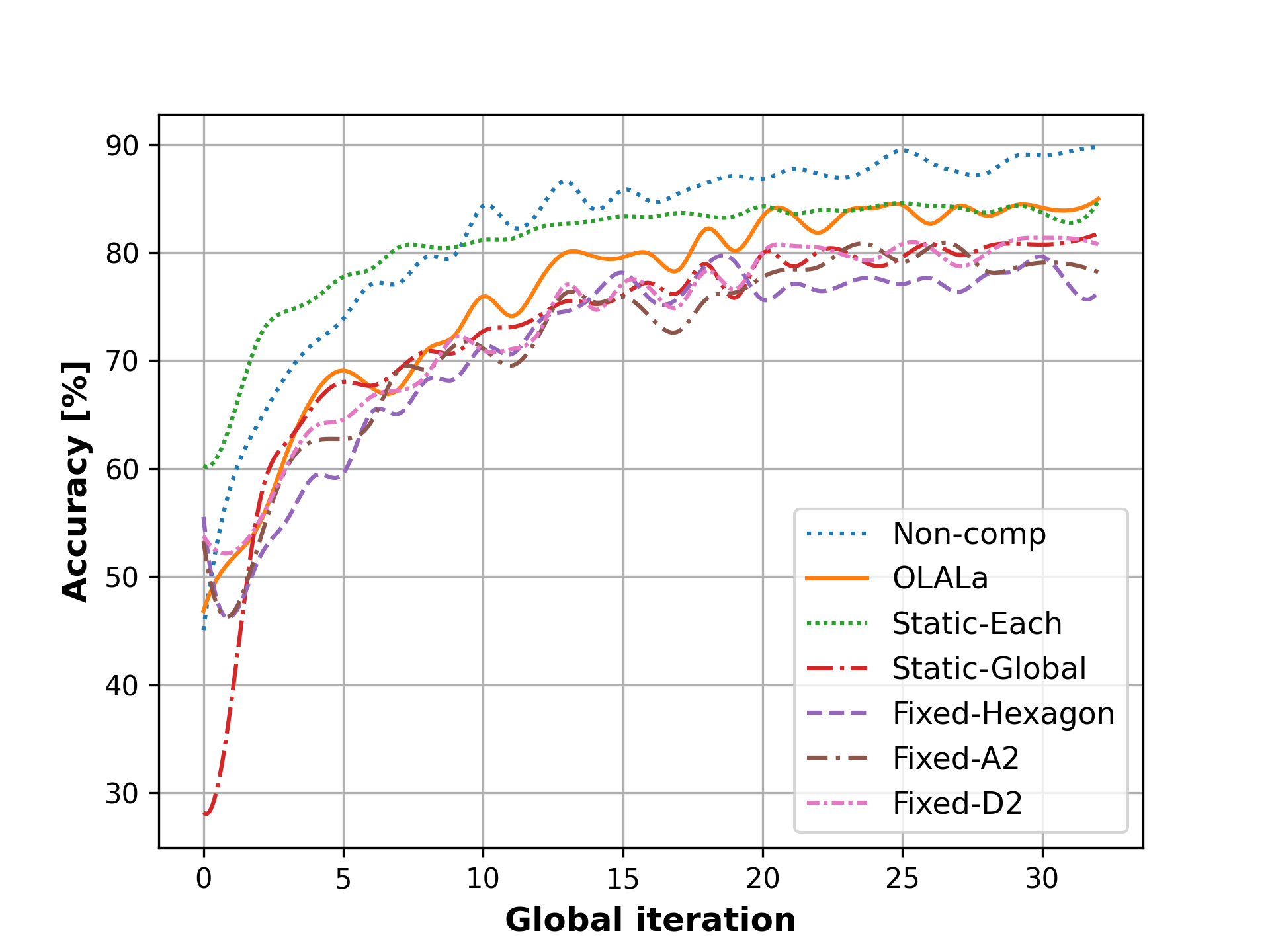}
    \caption{Accuracy vs. training rounds, MNIST, MLP, $R=3.5$.}
    \label{fig:mnistMLP}
\end{figure}

\begin{figure}
    \centering
    \includegraphics[width=\linewidth]{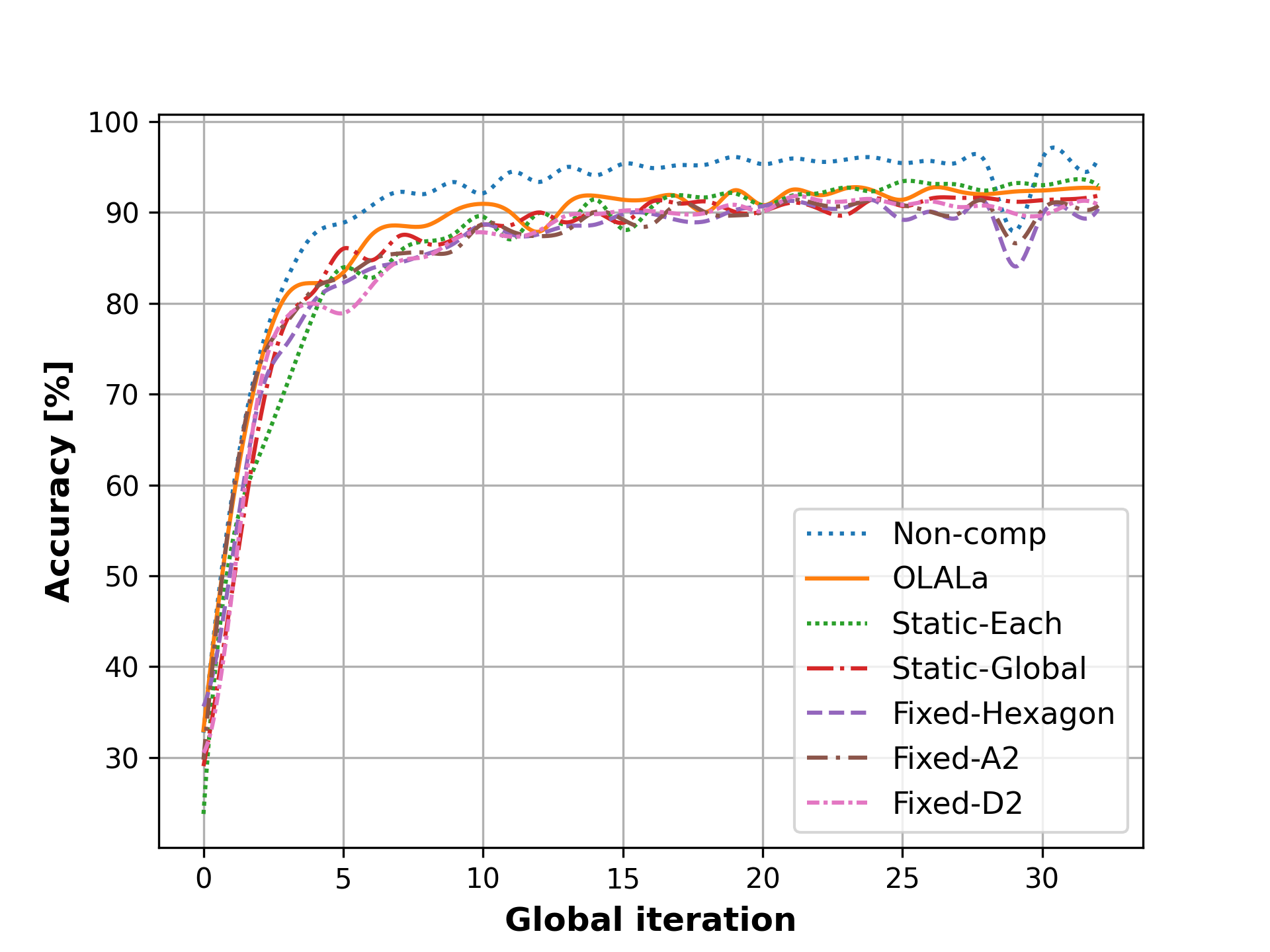}
    \caption{Accuracy vs. training rounds, MNIST, CNN, $R=3$.}
    \label{fig:mnistCNN}
\end{figure}

\begin{figure}
    \centering
    \includegraphics[width=\linewidth]{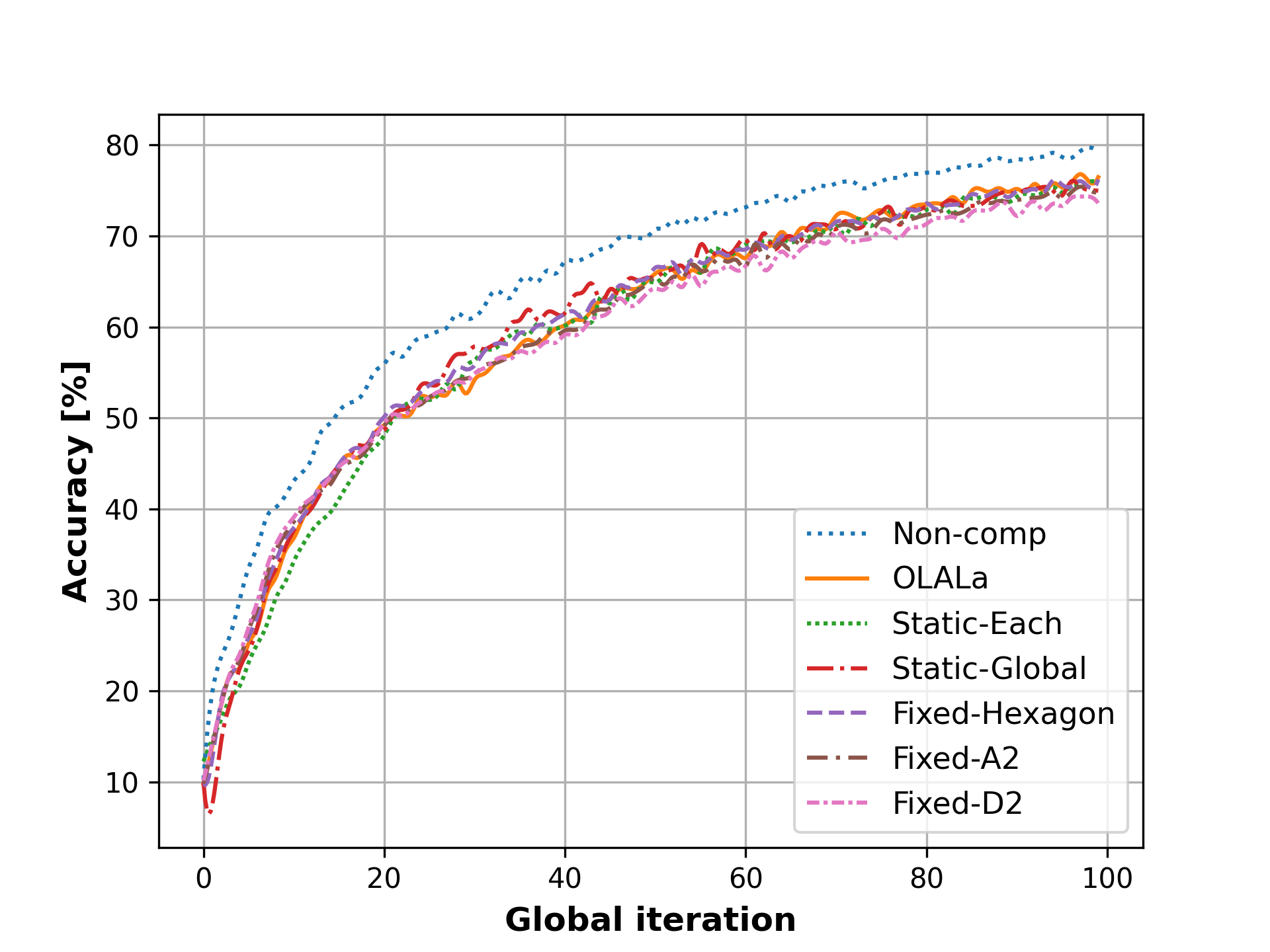}
    \caption{Accuracy vs. training rounds, CIFAR-10.  CNN, $R=3$}
    \label{fig:cifar}
\end{figure}

\begin{figure}
    \centering
    \includegraphics[width=\linewidth]{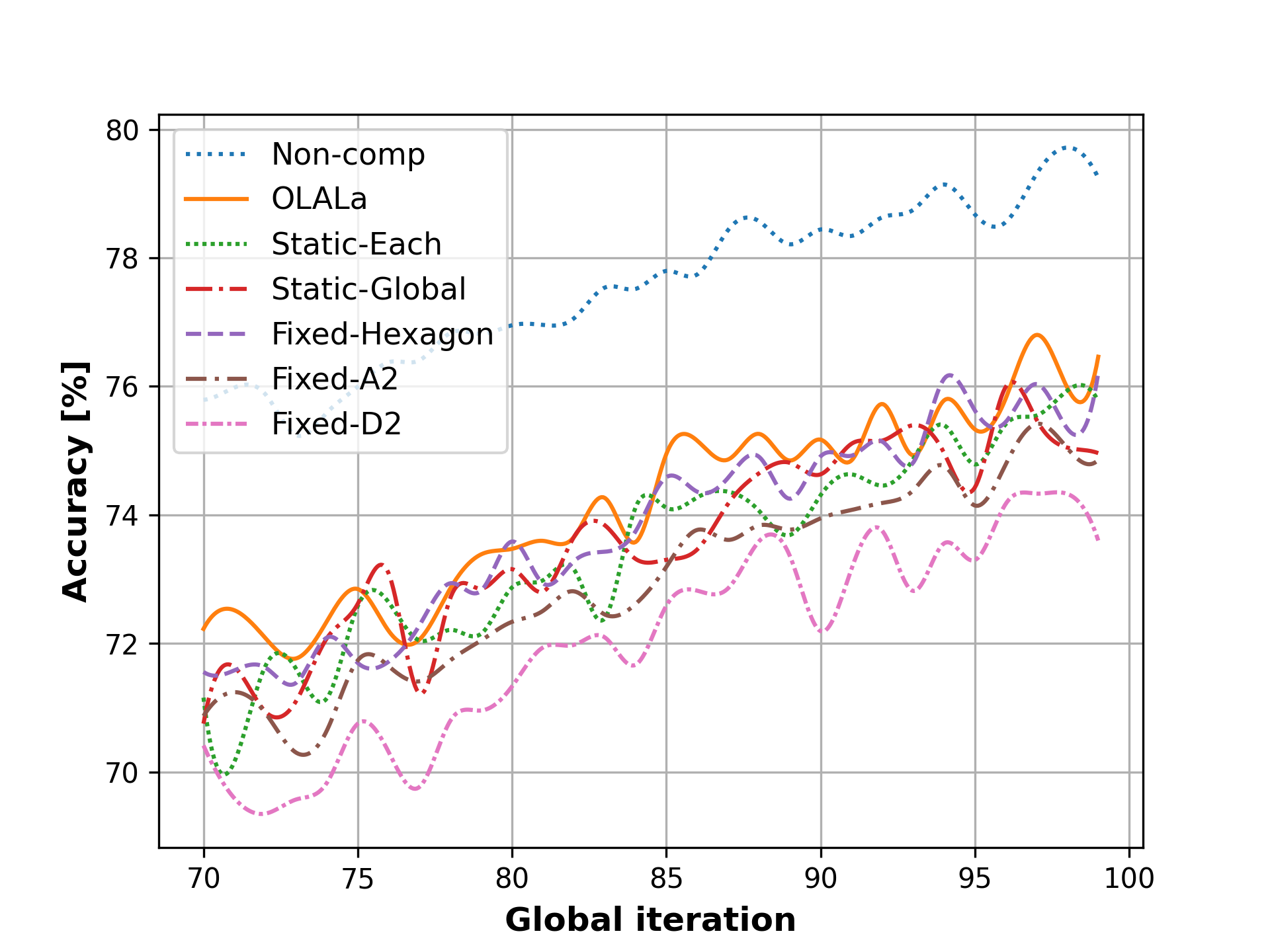}
    \caption{Accuracy vs. training rounds, CIFAR-10.  CNN, last 30 rounds, $R=3$}
    \label{fig:cifar2}
\end{figure}

{\bf Training Dynamics}:
To further compare the behavior of different quantization schemes, we examine the learning curves, i.e., the evolution of validation accuracy over training rounds. We report the results  three representative settings. 
The learning curve achieved for MNIST with the linear model and MLP (both at rate $R=3.5$ bits per weight) are reported in Figs.~\ref{fig:mnistLinear}-\ref{fig:mnistMLP}, respectively, while Fig.~\ref{fig:mnistCNN} shows the corresponding learning curve achieved with the CNN model (at a slightly lower rate $R=3$ where differences are better visible as follows from Table~\ref{tab:accVsCodewords}). 
In addition, we report the learning curves achieved for CIFAR-10 with CNN in Fig.~\ref{fig:cifar}, where we magnify the last 30 rounds for visibility in Fig.~\ref{fig:cifar2}. There, we considered a higher learning rate of $R=5$ bits per weight.


Across all configurations, we consistently observe the following trends: 
$(i)$ OLALa outperforms all other schemes in both convergence speed and final accuracy, especially as model complexity increases. This validates the value of round-wise adaptation to model update statistics; 
$(ii)$ Using the learning framework of \ac{name} in a static manner can also be beneficial. Specifically,  the per-user static strategy ("Static-Each") performs closely to OLALa in early training rounds, but its lack of temporal adaptivity limits its effectiveness when update distributions change over time; 
$(iii)$ Learning a single shared lattice ("Static-Global") consistently outperforms fixed lattice baselines but lags behind personalized schemes, indicating the importance of user-specific adaptation in heterogeneous settings;
$(iv)$ Fixed lattices (e.g., Hexagon, $A_2$, $D_2$) achieve lower performance overall and exhibit slower convergence, despite often having higher average SNRs, highlighting underscoring the need for task- and user-aware lattice learning.

\section{Conclusions}
\label{sec:conclusions}
We introduced  an online learned adaptive lattice quantization framework for \ac{fl}. Based on a theoretical analysis that highlights the potential benefits of using adaptive lattices, we proposed \ac{name}, which enables each client to learn and adapt a low-complexity lattice quantizer during training, allowing the quantization scheme to track the evolving distribution of local model updates. Our experimental study demonstrated that \ac{name} consistently outperforms fixed and statically learned lattice quantizers in terms of both convergence and final accuracy, offering a principled, efficient, and theoretically sound approach for communication-aware \ac{fl} in heterogeneous environments.


\begin{appendix}
\numberwithin{lemma}{subsection} 
\numberwithin{corollary}{subsection} 
\numberwithin{remark}{subsection} 
\numberwithin{equation}{subsection}	
This appendix characterizes, under assumptions \ref{itm:heterogenity}-\ref{itm:obj_smooth_convex}, the distortion induced by \ac{name}-aided \ac{fl}, as well as its convergence properties to  $\vw_{\rm opt}$ for growing number of global rounds $T$. These results follow the proof steps of \cite[Thm. 1]{li2019convergence}.

It it  noted that while analyzing $\Tilde\vw_{t+1}$ in~\eqref{eq:compressed_FedAvg_update_rule}, there are two different sources of randomness involved: 
a) stochastic gradients; and
b) probabilistic quantization.
As this pair is mutually independent with respect to both $u$ (user index) and $t$ (global \ac{fl} iteration index); the low of total expectation can be employed in order to isolate each stochastic term. Equivalently, each of which can be distinguished using the notation $\E_{\{\cdot\}}(\cdot)$. Yet, this subscript is dropped in the sequel for ease of notation.

\subsection{Additional Notations} 
For convenience, we define the averaged `full', stochastic, and compressed stochastic gradient, respectively, as follows:
\begin{align}
\vg_t &\triangleq \frac{1}{U}\sum_{u=1}^U \nabla F_u(\Tilde\vw_t) \label{eq:full_avg_grad_def}\\ 
\widehat{\vg}_t &\triangleq \frac{1}{U} \sum_{u=1}^U \nabla F_u(\Tilde\vw_t, \sample) \label{eq:rand_avg_grad_def}\\
\widehat{\vg}^{\rm SDQ}_t &\triangleq \frac{1}{U} \sum_{u=1}^U Q^{\rm SDQ}_{\cL_{\gamma^t_u}(\vG^t_u)} \left(\nabla F_u(\Tilde\vw_t, \sample)\right)
\label{eq:compressed_rand_avg_grad_def}.
\end{align}
According to \ref{itm:heterogenity}, the stochastic gradient is an unbiased estimator of the `full' one, i.e., 
\begin{equation}\label{eq:stoch_avg_grad}
    \E[\widehat{\vg}_t] = \vg_t. 
\end{equation}
Additionally, by \ref{itm:zero_overeloading} and \eqref{eq:sdq_error}-\eqref{eq:sdq_moments}, we obtain
\begin{equation}\label{eq:compress_stoch_avg_grad}
    \E[\widehat{\vg}^{\rm SDQ}_t]  
    =
    \E\left[\widehat{\vg}_t\right] + 
    \frac{1}{U} \sum_{u=1}^U 
    \E\left[\ve^{\rm SDQ}_{t,u}\right]
    = \vg_t.
\end{equation}

\subsection{Proof of Theorem \ref{thm:bounded_var}}\label{app:bounded_var_proof}
Using the auxiliary definitions of \eqref{eq:full_avg_grad_def}-\eqref{eq:compressed_rand_avg_grad_def}; and Theorem~\ref{thm:SDQ} given \ref{itm:zero_overeloading}, it holds that
    \begin{align*}
        &\E\left[\left\|\widehat{\vg}^{\rm SDQ}_t-\widehat{\vg}_t+\widehat{\vg}_t - \vg_t \right\|^2\right] \notag \\
        &= 
        \E\left[\left\|\widehat{\vg}^{\rm SDQ}_t-\widehat{\vg}_t \right\|^2\right] + 
        \E\left[\left\|\widehat{\vg}_t - \vg_t \right\|^2\right]
        \\  
        &\overset{}{=}  
       \frac{1}{U^2}\sum_{u=1}^{U} 
       \E\left[
       \left\|\ve^{\rm SDQ}_{t,u}\right\|^2 
       +
       \left\|  \nabla F_u(\Tilde\vw_t, \sample) - 
       \nabla F_u(\Tilde\vw_t)\right\|^2 \right] \\
       &\overset{\eqref{itm:bounded_variance}}{\leq} 
        \frac{1}{U^2}\sum_{u=1}^{U} 
       \left(
       \sigma^2_{\rm SDQ}\left(\cL_{\gamma^t_u}(\vG^t_u)\right)
       +
       \sigma^2_u
       \right).
    \end{align*}

\subsection{Proof of Theorem \ref{thm:FL Convergence}}\label{app:FL Convergence_proof}
We being by stating the following lemma, and deferred its proof to Section~\ref{ssec:one_step_sgd_proof}.

\begin{lemma}[Result of one-step SGD.]\label{lemma:one_step_sgd}   
Assume Assumption \ref{itm:obj_smooth_convex} hold. If $\eta_t \leq \frac{1}{2L}$, then 
\begin{align*}
    \E \left[ \| \Tilde\vw_{t+1} - \vw_{\rm opt} \|^2 \right] \leq (1 - \eta_t \mu) \E \left[ \| \Tilde\vw_t - \vw_{\rm opt} \|^2 \right] \\ 
    + \eta_t^2 \E \left[ \| \vg_t - \widehat{\vg}^{\rm SDQ}_t \|^2 \right] 
    + 2L \eta_t^2 \Gamma.
\end{align*}
\end{lemma}

Now, let $\Delta_t = \E\left\|\tilde\vw_t - \vw_{\rm opt}\|^2\right]$. From Lemma \ref{lemma:one_step_sgd} and Theorem~\ref{thm:bounded_var}, it follows that
\begin{align*}
    \Delta_{t+1} \leq (1 - \eta_t \mu)\Delta_t + \eta_t^2 B_t,\\ 
    B_t = \frac{1}{U^2}\sum_{u=1}^{U}
    \left(\sigma^2_u + \sigma^2_{\rm SDQ}\left(\cL_{\gamma^t_u}(\vG^t_u)\right) \right)+ 2L\Gamma.
\end{align*}

For a diminishing step-size, we set 
$\eta_t = \frac{\beta}{t + \nu}$ for some 
$\beta > \frac{1}{\mu}$ and $\nu > 0$ such that $\eta_1 \leq \frac{1}{2L}$. Then, 
in \cite[Thm. 1]{li2019convergence}, it is proved by induction that 
\[
\Delta_t \leq \frac{\textcolor{black}{v_t}}{\nu + t},\quad
\textcolor{black}{v_t} = \max\left\{\frac{\beta^2}{\beta\mu - 1}\textcolor{black}{\max_{t'=1,\dots,t}B_{t'}}, (\nu + 1)\Delta_1\right\}.
\]

Following \ref{itm:obj_smooth_convex}, $F(\cdot)$ is $L$-smoothness, and therefore
\[
E[F(\tilde\vw_t)] - F(\vw_{\rm opt}) \leq \frac{L}{2}\Delta_t \leq \frac{L}{2}\frac{\textcolor{black}{v_t}}{\nu + t}.
\]

Specifically, if we choose $\beta = 2/\mu$, $\nu + 1 = \max\left\{8L/\mu, 1\right\}$, and denote $\kappa = L/\mu$, then
$\eta_t = 2/\left(\mu(\nu + t)\right)$.
Finally, we have
\begin{align*}
    \textcolor{black}{v_t} &= \max\left\{
    \frac{\beta^2}{\beta\mu - 1}\textcolor{black}{\max_{t'=1,\dots,t}B_{t'}}, (\nu + 1)\Delta_1\right\} \\
   & \leq 
    \frac{\beta^2 }{\beta\mu - 1}\textcolor{black}{\max_{t'=1,\dots,t}B_{t'}} + (\nu + 1)\Delta_1 \\ 
    &\leq 
    \frac{4}{\mu^2}\textcolor{black}{\max_{t'=1,\dots,t}B_{t'}} + (\nu + 1)\Delta_1;
\end{align*}
and consequently,
\begin{align*}
    &\E[F(\tilde\vw_t)] - F(\vw_{\rm opt}) 
    \leq \frac{L}{2}\frac{\textcolor{black}{v_t}}{\nu + t} \\
    &\qquad\leq \frac{\kappa}{\nu + t}\left(\frac{2}{\mu}\textcolor{black}{\max_{t'=1,\dots,t}B_{t'}} + \frac{\mu(\nu + 1)}{2}\Delta_1\right),
\end{align*}
concluding the proof.

\subsection{Proof of Theorem~\ref{thm:gamma_dependent_optimal_G}}
\label{app:gamma_dependent_optimal_G_proof}
To prove Theorem~\ref{thm:gamma_dependent_optimal_G}, we begin by expressing the distortion associated with \ac{sdq} using the following  identity \cite[Eq. 6-7]{zamir1996lattice}:
\begin{equation*}
\sigma^2_{\rm SDQ}(\cL_\gamma(\vG)) = \cG(\vG) \cdot \det(\vG)^{2/L};\
\cG(\vG)\triangleq
\frac{1}{L} \cdot \frac{\int_{\cP_0} \|\vx\|^2 \, d\vx}{\det(\vG)^{1+2/L}}
,
\end{equation*}
where $\cP_0$ is the basic lattice cell (see Section~\ref{subsec:lattice_coding}), and $\cG(\vG)$ is coined the {\em normalized second-order moment} of the lattice.

Now, parameterize the generator using a scale factor $a$ and a unit-volume shape matrix $\vA \in \mathbb{R}^{L \times L}$ ($\det(\vA) = 1$) via $\vG = a \vA$.
Thus, $\det(\vG) = a^L$, and
\begin{equation}\label{eq:distortion_term}
 \quad \sigma^2_{\rm SDQ}(\cL_\gamma(\vG)) = \cG(\vA) \cdot a^2,
\end{equation}
where, since $\cG(\vG)$ is invariant to scale, $\cG(\vG)=\cG(\vA)$.

According to Theorem~\ref{thm:gamma_dependent_optimal_G}, the quantization rate is fixed to $R$, setting the codebook size as $2^{LR}$. Consequently, following the fundamental definition of a (truncated) lattice, it holds that
\begin{align}\label{eq:changing_variables}
2^{LR} = |\cL_\gamma(a \vA)| = \left| \left\{ \vz \in a \vA \mathbb{Z}^L : \|\vz\| \leq \gamma \right\} \right| \notag\\
\overset{\vz = a \vA \vz'}{=}\left| \left\{ \vz' \in \mathbb{Z}^L : \|\vA \vz'\| \leq \frac{\gamma}{a} \right\} \right|.
\end{align}

Next, define $r_{\vA}(R)$ as the {\em minimal} radius for which the count in \eqref{eq:changing_variables} equals $2^{LR}$, i.e.,
\begin{align}\label{eq:minimal_radius}
r_{\vA}(R) = \argmin_{r\in{\R}^{+}} \Big\{\left| \left\{ \vz' \in \mathbb{Z}^L : \|\vA \vz'\| \leq r \right\} \right| = 2^{LR}\Big\},
\end{align}
As a result, from \eqref{eq:changing_variables}-\eqref{eq:minimal_radius}, it follows that the most restricted configuration for $a$, in the presence of a fixed $R$ and $\vA$, is $a = \gamma / r_{\vA}(R)$. 
Then, substituting it back into the distortion term in \eqref{eq:distortion_term}, yields:
\begin{align}\label{eq:minimal_radius}
\sigma^2_{\rm SDQ}(\cL_\gamma(\vG)) = \cG(\vA) \cdot \left( \frac{\gamma}{r_{\vA}(R)} \right)^2.
\end{align}
Using \eqref{eq:minimal_radius}, we can define the following optimization problem:
\begin{equation}\label{eq:distortion_optim} 
\argmin_{\vG} 
\sigma^2_{\rm SDQ}(\cL_\gamma(\vG))
=
\gamma^2\cdot \argmin_{\vA \in \mathrm{SL}_L(\mathbb{R})} \frac{\cG(\vA)}{\left(r_{\vA}(R)\right)^2} 
,
\end{equation}
where $\mathrm{SL}_L(\mathbb{R})$ is the set of real $L \times L$ matrices with  $\det(\cdot) = 1$; concluding the proof.

\subsection{Deferred Proof of Lemma~\ref{lemma:one_step_sgd}}
\label{ssec:one_step_sgd_proof}
Using \eqref{eq:compressed_FedAvg_update_rule} and the auxiliary definitions of \eqref{eq:full_avg_grad_def},\eqref{eq:compressed_rand_avg_grad_def}; we have 
\begin{align*}
   & \E \left[ \| \Tilde\vw_{t+1} - \vw_{\rm opt} \|^2 \right]
    \\
    &= \E \left[ \| \Tilde\vw_t - \vw_{\rm opt} -\eta_t\vg_t 
    + \eta_t\vg_t - \eta_t\widehat{\vg}^{\rm SDQ}_t \|^2 \right]    \\
     &=\E \left[ \| \Tilde\vw_t - \vw_{\rm opt} -\eta_t\vg_t \|^2\right]    
   + \eta^2_t\E \left[ \| \vg_t - \widehat{\vg}^{\rm SDQ}_t \|^2 \right]    \\
   &\quad +2\eta_t\E  \left[\left\langle \Tilde\vw_t - \vw_{\rm opt} -\eta_t\vg_t, \vg_t - \widehat{\vg}^{\rm SDQ}_t \right\rangle  \right].
\end{align*}
Note that the last term is zero according to \eqref{eq:compress_stoch_avg_grad}, while the second term is bounded in Lemma~\ref{thm:bounded_var}.
As for the first summand, it is splitted into three terms:
\begin{align}\label{eq:A_term_bound}
A &\triangleq \|\Tilde\vw_t - \vw_{\rm opt} - \eta_t \vg_t\|^2 \notag\\ 
&=\|\Tilde\vw_t - \vw_{\rm opt}\|^2 - \underbrace{2\eta_t \langle \Tilde\vw_t - 
\vw_{\rm opt}, \vg_t \rangle}_{A_1}
+ \underbrace{\eta_t^2 \|\vg_t\|^2}_{A_2}.
\end{align}

From the L-smoothness of $F_u(\cdot)$ in \ref{itm:obj_smooth_convex}, and noting $F_u^\star=\min_{\vw} F_u(\vw)$, it follows that
\begin{equation}\label{eq:L_smoohth_inequality}
\|\nabla F_u(\Tilde\vw_t)\|^2 \leq 2L \left(F_u(\Tilde\vw_t) - F_u^\star\right);    
\end{equation}
that once combined with the convexity of $\|\cdot\|^2$, implies that:
\begin{align}\label{eq:A2_term_bound}
A_2 \leq \eta_t^2 \sum_{u=1}^U \frac{1}{U} \left\|\nabla F_u(\Tilde\vw_t)\right\|^2 \leq \frac{2L\eta_t^2}{U} \sum_{u=1}^U  \left(F_u(\Tilde\vw_t) - F_u^\star\right).
\end{align}
Note that
\begin{align}\label{eq:A1_term_bound}
A_1 &= -2\eta_t \langle \Tilde\vw_t - \vw_{\rm opt}, \vg_t \rangle \notag\\
&=-2\eta_t \left\langle \Tilde\vw_t  - \vw_{\rm opt}, \sum_{u=1}^U \frac{1}{U} \nabla F_u(\Tilde\vw_t) \right\rangle.
\end{align}

By the $\mu$-strong convexity of $F_u(\cdot)$ in \ref{itm:obj_smooth_convex}:
\begin{align*}
&- \langle \Tilde\vw_t - \vw_{\rm opt}, \nabla F_u(\Tilde\vw_t) \rangle \\
&\leq -\left(F_u(\Tilde\vw_t) - F_u(\vw_{\rm opt})\right) - \frac{\mu}{2} \|\Tilde\vw_t - \vw_{\rm opt}\|^2.
\end{align*}

Combining \eqref{eq:A1_term_bound} and \eqref{eq:A2_term_bound} bounds into \eqref{eq:A_term_bound} results with
\begin{align*}
&A \leq 
\|\Tilde\vw_t - \vw_{\rm opt}\|^2 + \frac{2L\eta_t^2}{U} \sum_{u=1}^U \left(F_u(\Tilde\vw_t) - F_u^\star\right) \\
&\quad- 2\eta_t \sum_{u=1}^U \frac{1}{U}   
\left(F_u(\Tilde\vw_t) - F_u(\vw_{\rm opt}) + \frac{\mu}{2} \|\Tilde\vw_t - \vw_{\rm opt}\|^2 \right)  \\
&=\|\Tilde\vw_t - \vw_{\rm opt}\|^2 + \frac{2L\eta_t^2}{U} \sum_{u=1}^U \left(F_u(\Tilde\vw_t) - F_u^\star\right) \\
&\quad- 2\eta_t \sum_{u=1}^U \frac{1}{U} \left( F_u(\Tilde\vw_t) - F_u(\vw_{\rm opt})\right)  
-\mu \eta_t \|\Tilde\vw_t - \vw_{\rm opt}\|^2. 
\end{align*}

Thus,
\begin{align}\label{eq:A_middle_bound}
&A \leq (1 - \mu\eta_t)\|\Tilde\vw_t - \vw_{\rm opt}\|^2 \notag\\
&+ \underbrace{\frac{2L\eta_t^2}{U} \sum_{u=1}^U  \left(F_u(\Tilde\vw_t) - F_u^\star\right)
- \frac{2\eta_t}{U} \sum_{u=1}^U  \left(F_u(\Tilde\vw_t) - F_u(\vw_{\rm opt})\right)}_{\triangleq A_3}.
\end{align}

To bound $A_3$, recall that by it holds that \eqref{eq:optimal_fl_model}
$F(\vw_{\rm opt})=\sum_{u=1}^U F_u(\vw_{\rm opt})/U$, and consequently
\begin{align}\label{eq:A3_term_bound}
    &A_3 \overset{(a)}{=} 
    \frac{-\overbrace{2\eta_t(1-L\eta_t)}^{\triangleq\gamma_t}}{U}\sum_{u=1}^U \left(F_u(\Tilde\vw_t)-F_u^\star\right)
    \notag\\
    &+ \frac{2\eta_t}{U} \sum_{u=1}^U \left(F(\vw_{\rm opt}) - F_u^\star \right)  
    \overset{(b)}{=} 
    \frac{-\gamma_t}{U}\sum_{u=1}^U \left(F_u(\Tilde\vw_t) - F(\vw_{\rm opt}) \right) 
    \notag\\
    &+ \frac{2\eta_t-\gamma_t}{U} \sum_{u=1}^U \left(F(\vw_{\rm opt}) - F_u^\star \right) 
    = -\gamma_t \underbrace{\left(F(\Tilde\vw_t) - F(\vw_{\rm opt}) \right)}_{\geq 0}
    \notag\\
    &+ 2L\eta^2_t\left(F(\vw_{\rm opt}) - \sum_{u=1}^U\frac{1}{U}F_u^\star \right) 
    \leq 2L\eta_t^2\Gamma.
\end{align}
where $(a)$ follows by adding and subtracting the term $\frac{2\eta_t}{U} \sum_{u=1}^U F_u^\star$; $(b)$ holds similarly using the term $F(\vw_{\rm opt})$; and the last inequity is true since it is given that $\eta_t\leq 1/(2L)$ what implies that $\gamma_t\geq 0$.
Plugging in \eqref{eq:A3_term_bound} into \eqref{eq:A_middle_bound}, we obtain:
\begin{align*}
A \leq (1 - \mu\eta_t)\|\Tilde\vw_t - \vw_{\rm opt}\|^2 + 2L\eta_t^2 \Gamma,
\end{align*}
and taking expectation on both sides completes the proof.

\end{appendix}

\bibliographystyle{IEEEtran}
\bibliography{IEEEabrv,bib}

\end{document}